\documentclass[runningheads]{llncs}

\usepackage[english]{babel}
\usepackage{rotating}
\usepackage{subfloat}
\usepackage{blindtext}
\usepackage{booktabs}
\usepackage{pdflscape}
\usepackage{threeparttable}
\usepackage{xcolor}
\usepackage{comment}
\usepackage{balance}
\usepackage{amsmath}
\usepackage[normalem]{ulem}

\AtBeginDocument{%
  \providecommand\BibTeX{{%
    \normalfont B\kern-0.5em{\scshape i\kern-0.25em b}\kern-0.8em\TeX}}}

\usepackage{hyperref}

\usepackage{array}
\newcolumntype{P}[1]{>{\centering\arraybackslash}p{#1}}

\providecommand{\ie}{\emph{i.e.,} }
\providecommand{\eg}{\emph{e.g.,} }

\providecommand{\ione}{\emph{(i)} }
\providecommand{\itwo}{\emph{(ii)} }
\providecommand{\ithree}{\emph{(iii)} }

\usepackage{subcaption}

\makeatletter
\AtBeginDocument{%
        \renewcommand{\@listi}
          {\setlength{\leftmargin}{\leftmargini}
           \setlength{\topsep} {1pt}
           \setlength{\parsep} {\parskip}
           \setlength{\itemsep}{1pt}}
         \renewcommand{\@listii}
          {\setlength{\leftmargin}{\leftmarginii}
           \setlength{\labelwidth}{\leftmarginii}
           \addtolength{\labelwidth}{-\labelsep}}
        \renewcommand{\@listiii}
          {\setlength{\leftmargin}{\leftmarginiii}
           \setlength{\labelwidth}{\leftmarginiii}
           \addtolength{\labelwidth}{-\labelsep}}}

\makeatother

\begin{document}

\title{Quantifying Nations' Exposure to Traffic Observation and Selective Tampering}

\author{Alexander Gamero-Garrido\inst{1,2} \and
Esteban Carisimo\inst{3} \and
Shuai Hao\inst{4}
\and \\ Bradley Huffaker\inst{1}
\and Alex C. Snoeren\inst{6}
\and Alberto Dainotti\inst{1,5}}
\authorrunning{A. Gamero-Garrido et al.}
\institute{CAIDA, UC San Diego \and Northeastern University \and
Northwestern University \and Old Dominion University 
\and Georgia Institute of Technology \and UC San Diego}

\maketitle

\begin{abstract}
Almost all popular Internet services are hosted in a select set of
countries, forcing other nations to rely on international connectivity
to access them.  We identify nations where traffic towards a large
portion of the country is serviced by a small number of Autonomous
Systems, and, therefore, may be exposed to observation or
selective tampering by these ASes.
We introduce the Country-level Transit Influence (CTI) metric to quantify the significance of a given AS on the international
transit service of a particular country.
By studying the CTI values for the top ASes in each country, we find 
that 34 nations have transit ecosystems that render 
them particularly exposed, where a single AS is privy to 
traffic destined to over 40\% of their IP addresses.
In the nations where we are able to validate our findings with in-country operators, 
our top-five ASes are 90\% accurate on average.
In the countries we examine, CTI reveals two classes of networks
frequently play a particularly prominent role: submarine cable operators and state-owned ASes.

\vspace{-0.2cm}

\end{abstract}

\section{Introduction}
\label{sec:intro}
The goal of this study is to identify
instances where a significant fraction of 
a country's inbound international traffic is managed 
by a select few networks.
Such networks are in a position to observe and tamper with a nation's traffic, 
as could any third-parties who infiltrate them (\eg using a phishing attack
or a remote vulnerability exploitation). \color{black}
For instance, observation---of unencrypted traffic and metadata---may 
be performed by
domestic or foreign actors with the purpose of 
conducting surveillance or espionage, respectively. Conversely,
selective tampering---for instance, with individual network flows 
carrying popular-application traffic---has been reported by actors that are
both domestic (e.g., government censorship)
and foreign (e.g., dis-information campaigns).

Because actual traffic information is difficult to obtain at a global scale, 
we instead quantify the fraction of a country's IP addresses
exposed to tampering and observation by specific networks.  While all IP addresses are clearly not created equal, 
they facilitate an apples-to-apples comparison across nations,
and the ranking of networks influencing a particular country. \color{black}
Traffic towards any given IP address is frequently handled by
so-called transit networks, \ie
those who sell connectivity to the rest of the Internet to other, customer networks
for a fee; customers include consumer-serving access networks. %

These transit networks are often unknown and unaccountable to end
users. %
This opacity may allow both domestic and foreign 
actors to observe or tamper with traffic---capabilities 
we term \textit{transit influence}---without
facing
diplomatic or political backlash from governments, 
activists or consumer groups.
We aim to bring transparency to the public regarding 
oversized observation and
tampering capabilities granted to specific transit networks
in a large group of nations. 

In order to reveal these crucial, nation-level topological
features, we develop the country-level transit influence (CTI) metric.
CTI quantifies the transit
influence a particular network exerts on a nation's traffic.
Studying transit influence requires an analysis of 
the global routing ecosystem which enables networks
to exchange traffic between them.
We extract information from the Border Gateway Protocol (BGP), the central 
system by which networks exchange interconnection information.
CTI is based on an analysis of a large compendia of BGP data 
\cite{ripeprobes,routeviews} and includes both 
topological and geographic filters designed to facilitate inference despite incomplete and biased data \cite{Luckie2013,Hegemony,dhamdhereevo}. 

We apply CTI in countries that lack peering facilities
such as Internet exchange points (IXPs) at which access networks might
connect directly with networks of other nations.
In these \color{black}\textit{transit-dominant} \color{black}nations, transit networks---often a select few based in
geographically distant
countries~\cite{Bischof,fanoupaths,galperin2013connectivity,Roberts}---serve as
the dominant form of connectivity to
the global Internet. %
Moreover, the lack of internationally connected, 
\color{black} domestic co-location 
facilities places these nations at
further risk of exposure to observation and tampering because
popular content is generally hosted abroad~\cite{Cai2013AHF,Edmundson,Mbaye,Gupta,ShahRouting}.

We employ a two-stage approach based on
a comprehensive set of 
passive inference and active measurements. First,
we identify transit-dominant countries. 
\color{black}
Countries that are transit dominant may be more exposed to observation and 
tampering by transit providers than countries where peering agreements 
are prevalent: the latter can receive some traffic from other 
countries through such peering agreements and bypass transit providers. 
\color{black}
Second, we quantify the transit influence of the networks serving 
each country using the CTI methodology, the central contribution of this study. 
We validate our 
findings from both stages with in-country network operators at 123 ASes 
in 19 countries who each confirm that our results are consistent with their understanding
of their country's networks.
These discussions, and our 
analyses showing the metric's stability, \color{black}
lend confidence to our inferences despite the considerable
technical challenges
in this measurement space. 

In addition to releasing our code and data, 
\color{black} our contributions include:
\begin{enumerate}
\item A new Internet cartography metric that quantifies the transit
influence a particular network exerts on a nation's traffic:
the country-level transit influence (CTI) metric, which ranges over $[0, 1]$. %
\item We apply CTI to infer the most influential 
transit networks in 75 countries that rely primarily 
on transit for international connectivity. %
These countries have, in aggregate, $\approx$1 billion Internet users
(26\% of the world~\cite{Internet68:online}).
We find that many of these countries have
topologies exposing them to observation or tampering: 
in the median case, the most influential transit network
manages traffic towards 
35\% of the nation's IP addresses.
\item We
identify two 
classes of ASes that are frequently influential: those who operate submarine 
cables and companies owned by national governments. 
\end{enumerate}

\textit{Ethical disclaimer}. 
We acknowledge several ethical implications of our work.  %
Our mass (validation) survey of operators was classified as exempt by our IRB.
Our reporting of available paths to repressive countries might trigger
government intervention to remove such paths. Another potential issue
is the identification of networks %
that would yield the most expansive observation or tampering capabilities
in a country, which is
potentially useful information for a malicious actor. We believe most
governments and sophisticated attackers already have access to this
information,  and that our study may lead to mitigation of these concerning
topological features; thus, the benefits
significantly exceed the risk.

\textit{Roadmap.}
The remainder of this paper is organized as follows. We start in
\S\ref{sec:method} with a high-level overview of our methodology
before 
describing how we assign
nationality to prefixes, ASes, and BGP vantage points
(\S\ref{sec:geolocation}). 
We introduce the CTI metric in \S\ref{sec:ti}.  We apply
CTI in 75 countries where international connectivity is predominantly
transit and %
describe our findings in \S\ref{sec:all-results}.
Then, we discuss in detail how we identified the transit-dominant
countries (\S\ref{sec:transit}).
We present our validation with operators
and stability analyses in \S\ref{sec:validationmain}.
\S\ref{sec:limitations} discloses
some limitations of our study while \S\ref{sec:related} compares
with prior work. 
Due to space constraints, we include further details and
a flowchart summarizing our full methodology 
in the appendix. We release the CTI code and datasets at 
\url{https://github.com/CAIDA/mapkit-cti-code}.
\color{black}

\section{Approach Overview}
\label{sec:method}

Conceptually, international Internet traffic crosses a nation's border
at some physical location, likely along a link connecting two routers.
For our purposes, we are not interested in the physical topology, but
the logical one: in which autonomous system(s) does international
traffic enter a nation on its way to access networks in that country
(i.e., origin ASes).  Topologically, these ASes can have two different
types of relationship with the first domestic AS encountered: transit
(provider-to-customer or \textit{p2c}) or peering (peer-to-peer or
\textit{p2p}).
We focus on countries where international connectivity
is dominated by transit (p2c) interdomain relationships as they are
easier to identify from public data sources.

\textbf{High-level model}. 
We look for evidence of a country's exposure to
observation or selective tampering by specific networks. 
Studying this exposure 
requires a quantitative model of the
reliance of the country's access networks, in aggregate, on specific transit networks. 
The model must factor in %
the size of the address space originated by each AS with presence in the country.
Intuitively, the greater the share of a country's IP
addresses that are served by a particular transit AS, the higher the
potential exposure of the nation's inbound traffic to observation or tampering
by that AS. 
The model must then produce a country-level metric of exposure for each transit network
serving the nation.  
To that end, we determine the frequency 
at which transit networks appear on routes towards the country's IP addresses.

We start our model %
by building a graph where nodes are ASes
and edges are connections between them,
weighted by address space. 
Then, a metric of node prominence on said graph
provides a quantitative assessment of how frequently a (transit) node $AS_t$ is traversed 
when delivering traffic from any given node to edge (origin) nodes.
The higher the value of this metric for any $AS_t$ in a given country,
the more exposed the transit ecosystem is.
At one extreme (most exposed) are countries with a
single transit provider (\eg a legally-mandated monopoly) connecting
every network in the country to the rest of the Internet; at the other
end are countries with many transit providers, each delivering traffic
to a small fraction of the nation's IPs.
Note that we do not need complete visibility of the graph (\eg backup links)
to infer potential exposure to observation or tampering, as traffic
will likely flow through the links that are visible given capacity constraints
on long-haul (incl. international) links~\cite{Akella,Zeitoun,LiuJingang,IsYourPl50:online}.

Our technical approach to build this conceptual
model using real data uses as inputs a combination of
two types of measurements: \ione passive, to study AS-level
connectivity, and
\itwo active, to study %
transit dominance.

\textbf{AS-level connectivity.}
We rely on two major input sources: BGP paths and prefixes from RouteViews~\cite{routeviews} and RIPE RIS~\cite{ripe}, 
and AS relationship inferences from CAIDA.
\color{black}
We begin with the 848,242 IPv4 prefixes listed in CAIDA's 
Prefix-to-Autonomous System mappings derived from RouteViews~\cite{pfixtoas},
excluding the 6,861 (0.8\%) prefixes with (invalid) length greater than 24, and 
the 9,275 (1.1\%) originated by multiple ASes. %
We find those prefixes in the 
274,520,778 IPv4 AS-level paths
observed in BGP table dumps
gathered by AS-Rank~\cite{ASRankAb8:online} 
from RIPE/RouteViews \cite{routeviews}\cite{ripe} 
during the first five days of
March 2020. We consider the set of prefixes and the ASes that
originate them on each observed path in combination with the 377,879
inferred AS-level relationships published by
CAIDA~\cite{asrelationships}.\footnote{In the 75 countries where we study transit influence,
no path contained any of: unallocated
ASes, loops, poisoned paths (where a non-clique AS is present between
two clique ASes, clique being the AS-level core of the Internet
inferred by \cite{asrelationships}); additionally, all paths towards
these countries are seen at least once per day across all five days.}

\textbf{Transit dominance}. Because we are focused only
on countries where
transit---as opposed to peering---is the main form of trans-border
connectivity, we use active measurements to identify and exclude
nations with evidence of foreign peering, \ie where an AS that originates
addresses geolocated to the country establishes a peering agreement
with another AS primarily based in another country\footnote{This
  ``nationality'' assignment is described in
  Sec. \ref{sec:nationality}.}.
We conduct
a two-week-long active measurement campaign (see Sec.~\ref{sec:campaign}) in May 2020 to
determine which countries are transit dominant
based on the business relationship %
between the ``border'' ASes traversed by our probe packets 
while entering the country
(as inferred by BdrmapIT~\cite{bdrmapit}).

\section{Definitions of Nationality}
\label{sec:geolocation}

CTI hinges on the correct nationality assignment for IP address
prefixes and BGP monitors. ASes are also assigned 
a nationality in the transit-dominance analysis. %
Given the diverse set
of information available, we devise distinct methods for each.
(We include an analysis of CTI stability given an alternative
geolocation input in \S \ref{sec:validationmain}).
\color{black}
For our purposes, a country is one of the 193 United Nations member states,
either of its two permanent non-member observer states, or Antarctica.

\textbf{Address prefixes}.
We first geolocate each IP address in every observed BGP prefix to a
country using
Netacuity~\cite{netacuity}.
Then, on a country-by-country basis, we count how many addresses in
each prefix are geolocated to that country.  If the number is less
than 256 (a /24), we round up to 256.  If Netacuity does not place
any of a prefix's IP addresses in a country, we attempt to find
a delegation block from the March 2020 RIR delegation
files~\cite{delegation} that covers the entirety of the prefix. If
there is one we assign all of the delegated prefix's addresses to the
indicated country.  Hence, while Netacuity can place a prefix in
multiple countries, at most one country will receive addresses through
the RIR process, and only if it was not already associated with the
prefix through Netacuity.
Netacuity
accounts for 95.1\% of all prefix-to-country mappings,
while delegation-derived geolocation accounts for the rest.

A particularly pressing concern with geolocation is the 
correct assignment of IP addresses belonging to large transit
ASes with a presence in many countries. 
We compute the fraction of a country's address space that 
is originated by ASes that have at least two thirds of
their addresses in that country.
In the vast majority of countries, the 
address space is dominated by ASes that are primarily domestic.

\color{black}

\label{sec:inboundfilter}

\textbf{BGP monitors}. As our study is focused on measuring inbound country-level connectivity,
we seek to limit our analysis to paths going towards addresses in the
target country from a BGP monitor located outside that country.
Hence, we confirm the BGP monitor locations listed by
RouteViews~\cite{routeviewslist} and RIPE RIS~\cite{riperislist}
through a set of active measurements. The details of this
process are included in Appendix~\ref{app-bgp-mon}.

\textbf{Autonomous Systems}.
\label{sec:nationality}
Our transit dominance analysis relies on a concept of AS
nationality, which is based on IP geolocation of the AS' 
originated addresses; for
transit providers, we also include the IP addresses
originated by direct customers. %
\color{black}
We classify each autonomous system $AS$ operating in a country $C$ as
being \textit{domestic}, $AS \in \mathrm{dom}(C)$, when the AS has at
least two thirds of its addresses in the country, and \textit{foreign}
otherwise. The vast majority (97.4\%) of ASes are classified as
domestic in one country, with the remaining small fraction being
classified as foreign in every country.  In fact, 89.8\% of ASes have
all of their address in a single country, and 98.6\% have a strict
majority of addresses in one country.

\begin{figure}[t]
\begin{minipage}{0.62\linewidth}
    \centering
    \includegraphics[width=0.99\textwidth]{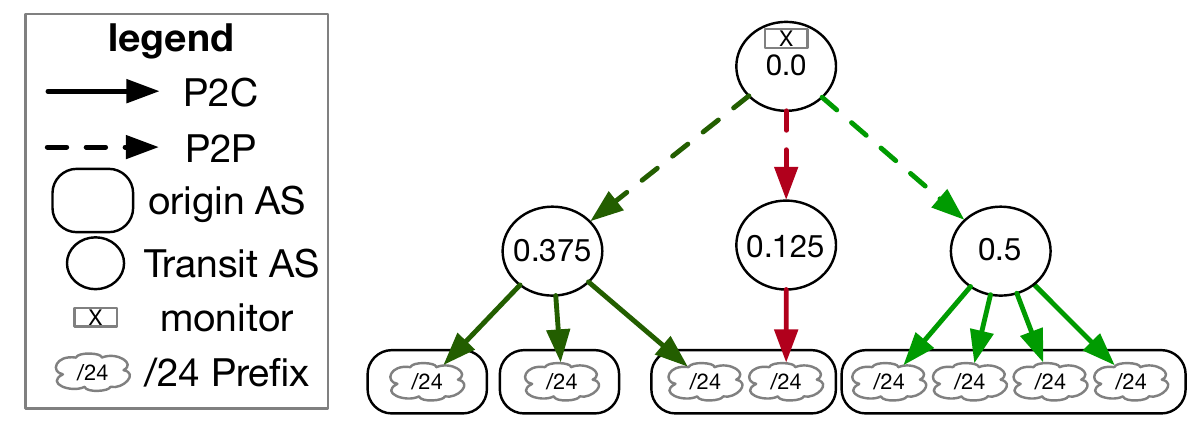}\\
    \caption{Example of Country-Level Transit Influence.}\label{fig:ativalues}
\end{minipage}\hfill
\begin{minipage}{0.36\linewidth}
    \centering
    \includegraphics[width=.99\textwidth]{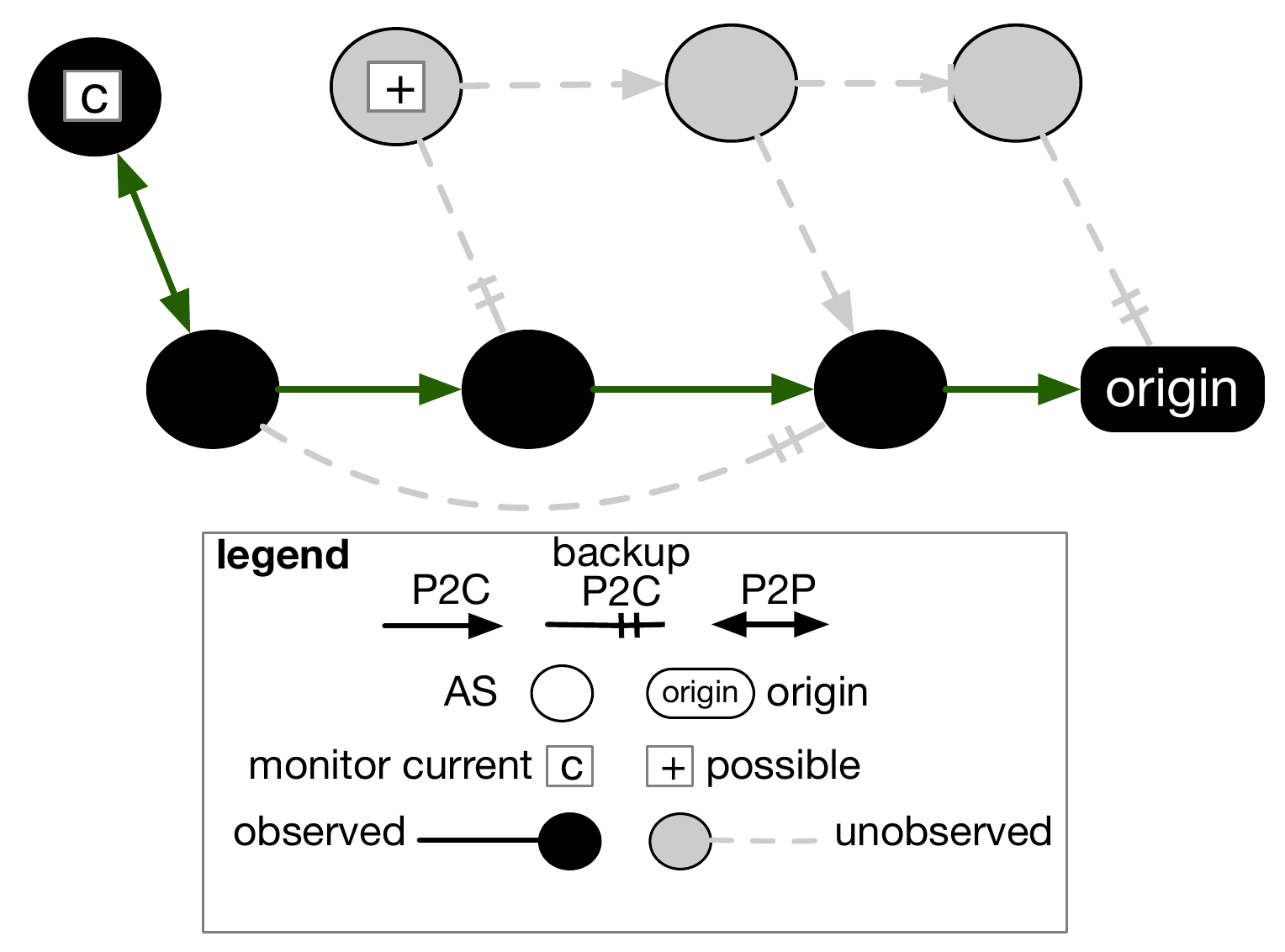}\\
    \caption{Unobserved paths in BGP.}\label{fig:backuplinks}
\end{minipage}\hfill
\end{figure}

\section{Transit Influence Metric}
\label{sec:ti} %

We define the transit influence $CTI_M(AS, C) \in [0,1]$
using a set of BGP monitors $M$ as
\begin{equation}
\label{eq:cti}
\sum_{m\in M} \left( \frac{w(m)}{|M|} \cdot  \sum_{{p} | \mathrm{onpath}(AS,m,p)} \left( \dfrac{a(p,C)}{A(C)} \cdot \dfrac{1}{d(AS,m,p)} \right) \right),
\end{equation}
\noindent where $w(m)$ is monitor $m$'s
weight (Sec.~\ref{sec:asdiversity}) among the set of monitors
(Sec.~\ref{sec:hegemony}); $\mathrm{onpath}(AS,m,p)$ is true if
$AS$ is present on a preferred path observed by monitor $m$ to a
prefix $p$, %
and $m$ is not contained within
$AS$ itself (Sec.~\ref{sec:p2cfilter});
$a(p,C)$ is the number of addresses in prefix $p$ geolocated to country $C$;
$A(C)$ is the total number of IP addresses geolocated to country $C$;
and $d(AS,p,m)$ is the number of AS-level hops between
$AS$ and prefix $p$ as viewed by monitor $m$ (Sec. \ref{sec:positionfilter}).

We illustrate CTI's use in Fig.~\ref{fig:ativalues}, with CTI
values for a toy example with three transit ASes and four origin ASes,
in a country with eight /24 prefixes: the transit AS on the right has
the highest CTI, since it serves the most addresses (half of the
country), followed by the transit AS on the left ($3/8$) and the AS in
the center ($1/8$). Note that the top AS has a CTI of 0, because it
hosts the BGP monitor from which the set of routes used in this
toy example are learned---hence, $\mathrm{onpath}(AS_t,m,p)$ is
always false for that AS.  Should that AS not be the host of
the BGP monitor (or be seen on these routes through another monitor),
it would have a CTI of 0.5---transit influence over the entire country
as an indirect transit provider (distance 2 from the prefixes).

Note that originating
addresses directly does not grant an AS transit influence, as
our focus is on identifying ASes that carry traffic to
destinations outside of their network.
 
\subsection{CTI components}

We explain the rationale for the various factors in
Eq.~\ref{eq:cti} in the following subsections. 

\subsubsection{Indirect transit discount.}
\label{sec:positionfilter}
As the number of AS-level hops from the origin increases,
so too does the likelihood that there exist alternative paths towards the same origin
AS of which we have no visibility (\eg backup links, less-preferred paths).
Fig. \ref{fig:backuplinks} shows this limitation in visibility for a toy example
with a single origin AS. There, given
the location of BGP monitor $C$ we see the AS-level chain in black, 
erroneously concluding that the origin AS has a single direct transit provider
and two indirect transit providers. In reality, there exists another set
of both direct and indirect transit providers (the AS-level chain in light
gray). We miss all these paths given that we do not have a 
monitor in any neighbor of a light-gray AS
(such as that marked with a plus sign). In this example we miss backup links of
the origin AS, as well as preferred links of the origin's direct transit provider,
and a backup link of both indirect transit providers.

As a coarse mechanism aimed at mitigating this limited visibility, we
discount the influence of transit providers in proportion to the
AS-level distance from the origin:  we apply a discount factor
as $1/1, 1/2,..., 1/k$, where
$k$ is the number of AS-level hops from the origin AS.
In practice, that means we do not discount the measurements of 
direct transit providers, as there the probability of missing
\color{black}
a backup or less-preferred link is lowest.
\color{black}
We note that this heuristic yields a conservative 
estimate of the observation opportunities of an indirect 
transit provider over traffic flowing towards a country.

\subsubsection{Prioritizing AS diversity.}
\label{sec:asdiversity} 
ASes can host more than one BGP monitor. In fact, more than 20 ASes in RIPE RIS
and RouteViews host multiple monitors; for instance,
AS3257-GTT hosts five.
In order to favor a topologically-diverse view (given the available
observations), if more than one monitor from the same AS sees an
announcement for the same prefix, we discount their %
observations to limit the influence of monitor ASes with multiple
monitors. Formally, the weight for each monitor $m$'s observation of a
prefix is $w(m) = 1/n$, where $n$ is the number of BGP
monitors in the AS that see an announcement of
that prefix.

\vspace{-0.2cm}
\subsection{Filtering ASes}

To correct for the limited, non-uniform coverage of the BGP
monitors that collect our table dumps, we apply a number of filters to
the set of paths over which we compute CTI.

\subsubsection{Provider-customer AS filter.}
\label{sec:p2cfilter}
BGP monitors by definition collect paths from the AS hosting the
monitor to the origin AS.  Therefore, we always exclude the AS hosting
the BGP monitor from the path to avoid inflating their transit
influence.
Further, we employ a heuristic that
attempts to consider only the portion of the path relevant to the
origin prefix, and ignore the portion dictated by the monitor's
topological location.

The intuition behind our filter is that, from the perspective of the
origin AS, there is a ``hill'' above it capped by the last observed
provider-customer (p2c, \ie transit) link, with traffic flowing from the hill's peak down
towards the origin.  The transit AS in that link is the highest point
in the path we want to keep, as it directs traffic towards its
customer (and its customer's customers, if applicable).  After
reaching that topological peak, we discard any other AS present
in the path.  The remaining path would then include the origin AS, its
direct or indirect transit provider at the topological peak, and any
other ASes appearing between the origin AS and the direct or indirect
transit provider. Note that this filter excludes
peers of the transit provider at the peak---appearing
between the topological peak and the AS hosting the BGP 
monitor---since we only apply
CTI in transit-dominant countries, and therefore
these peers are unlikely to be central to the country's 
connectivity. \color{black}

Formally, for the analysis presented in this paper, we refine
$\mathrm{onpath}(AS_t, m, p)$ to be true only if the path observed at
monitor $m$ has at least one inferred p2c link where the customer is
either the origin of $p$ or closer to it than $AS_t$, \ie we discard
paths where there
is no topological peak from the perspective of the origin.  This
heuristic discards 0.2\% of the paths observed by our monitors.  In
the median country we discard 0.2\% of paths using this filter, with
0.3\% being the average case.  In all countries we keep over 98.6\% of
paths.

This filter ensures that at
least one AS (the inferred customer of the transit AS) 
relies on at least one other AS (the
inferred transit provider) 
for transit from and towards the core of the Internet. As we
aim to measure transit influence, these business relationships are an
important source of information: merely being directly connected to an
AS path that reaches the origin AS in a given country does not necessarily make an AS
influential; being a direct provider of the origin, or of an AS closer
to the origin, lends more confidence to our inference of 
influence\footnote{\color{black}Refer to \cite{alexgdissertation}
(\S 2.1.5 and \S 4.2.4) for an extended discussion of the intuition behind the CTI model.\color{black}}.
\subsubsection{CTI outlier filtering.}
\label{sec:hegemony}
Finally, we filter BGP-monitor-location noise by removing
outlier estimates of transit influence---both overestimates and
underestimates resulting from the AS hosting a 
BGP monitor being topologically too
close or too far from the origin AS---to get an accurate assessment of
transit influence towards that origin.  We implement a filter recently
proposed for another AS-topology metric (AS
hegemony~\cite{Hegemony}, see \S\ref{sec:related}).
Specifically, we compute the CTI of each transit provider
$AS_t$ using BGP monitors from
each monitor-hosting $AS_h$ independently, as $CTI_{m(AS_h)}(AS_t,C)$, where $m(AS_h)$ is the set of
monitors within $AS_h$.
We determine which potentially-biased $AS_h$ have gathered
observations producing $CTI_{m(AS_h)}(AS_t,C)$ values in the bottom
and top 10\% of all values for that transit provider in that country
and
disregard all paths observed by monitors hosted in these potentially-biased $AS_h$.
As in~\cite{Hegemony}, we 
implement outlier filtering only where we have observations of
$CTI_{m(AS_h)}(AS_t,C)$ from 10 or more $AS_h$, which occurs for
58.4\% of transit AS-country pairs in our sample (a single AS can
operate in multiple countries).

\section{Country-Level Transit}%
\label{sec:all-results}

In this section we present the results of applying our CTI metric to the transit ecosystem of 75 countries with little-to-no international peering.  (We describe our method for selecting these countries in \S\ref{sec:transit}.)
We provide a high-level characterization of the transit ecosystem in
each country by comparing the CTI scores of the top-5 ASes ranked by
CTI (Sec.~\ref{sec:results:overview}), as well as a set of ASes
that appear in the top 5 of many countries (at least 10).  
Our hypothesis is that these countries show different transit
profiles as a consequence of the socioeconomic and geopolitical
diversity of the sample:
from high exposure to observation, where one AS is the most influential transit provider and
others are very marginal, to less exposed countries with
an ensemble of ASes with similar values of CTI. 

Investigating the companies operating the ASes with high CTI,
we find two prominent groups of organizations:
submarine cable operators (Sec.~\ref{sec:results:cable}) and
state-owned providers (Sec.~\ref{sec:stateowned}).  For the former,
their operation of physical infrastructure 
connected to the country may underpin their high transit influence.
\color{black} With regards to state-owned ASes, %
providing transit may give governments the ability
to expand their footprint beyond addresses they originate, \eg
through a state-owned broadband provider. In some cases, state
ownership of a transit provider may follow their investment in a
submarine cable or landing station, while in others it
may reflect the government's intention to enact censorship. 
We limit our analysis to the discovery of the
transit footprint of the state, without delving into the underlying
motives.

\subsection{CTI distribution across countries}
\label{sec:results:overview}

\begin{figure}
\vspace{-4mm}
    \centering
    \includegraphics[width=.92\textwidth]{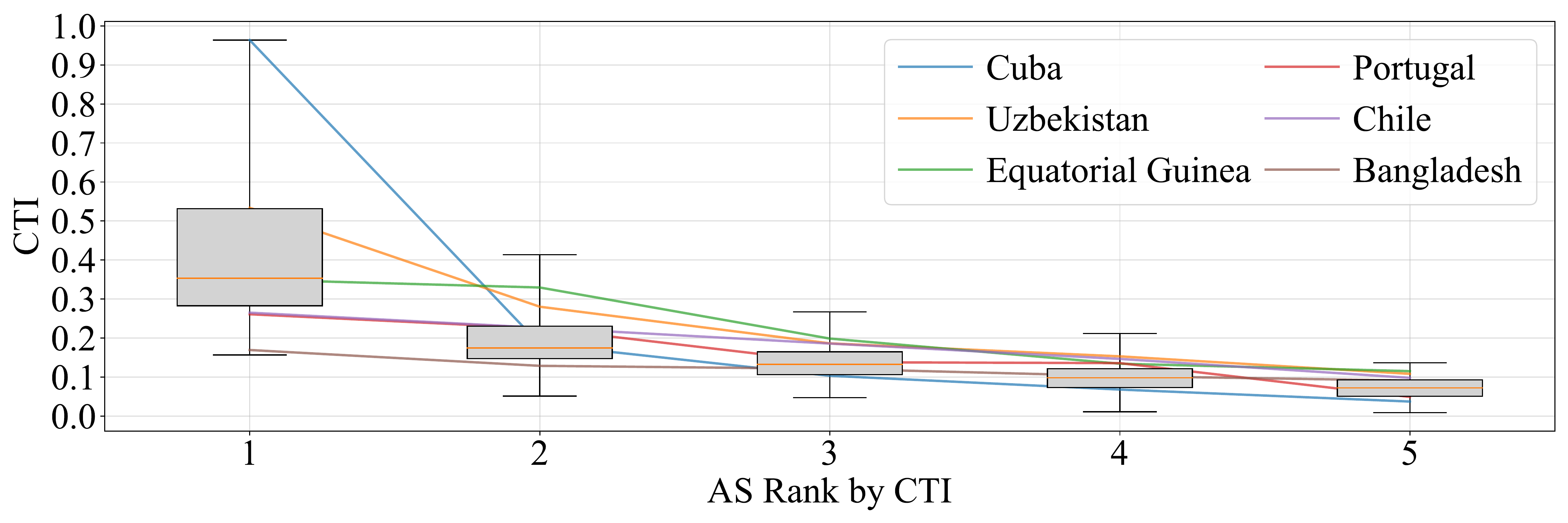}\\
    \caption{Boxplot of CTI distributions for the top-5 ASes in each country.}
\label{fig:ctioverview}
\vspace{-4mm}
\end{figure}

In this subsection we present an overview of the CTI distribution
across countries. Countries with a 
top-heavy distribution of CTI values are particularly 
exposed to specific networks. Other nations with a more
flat distribution signal an ecosystem that is less exposed
to prominent transit ASes.
Fig.~\ref{fig:ctioverview} shows the distribution of CTI values for
ASes ranked in the top 5 by CTI in each country.  
In 51 countries, the top-ranked AS has CTI $\geq$ 0.3,
signaling high exposure to observation and tampering by that specific network.

The distribution of CTI rapidly declines across AS
rank, with the median halving from the first to the second position.
In 54 countries, 
CTI declines by over 30\% from the top-ranked AS to its successor;
the average and median decline across all countries are 50\% and 47\%.
This suggests that in the vast majority of countries in our sample, a
single AS is particularly prominent in terms of its capabilities
to observe or tamper with traffic.

\subsubsection{Individual nations.}

\begin{figure}
    \centering
    \includegraphics[width=.92\textwidth]{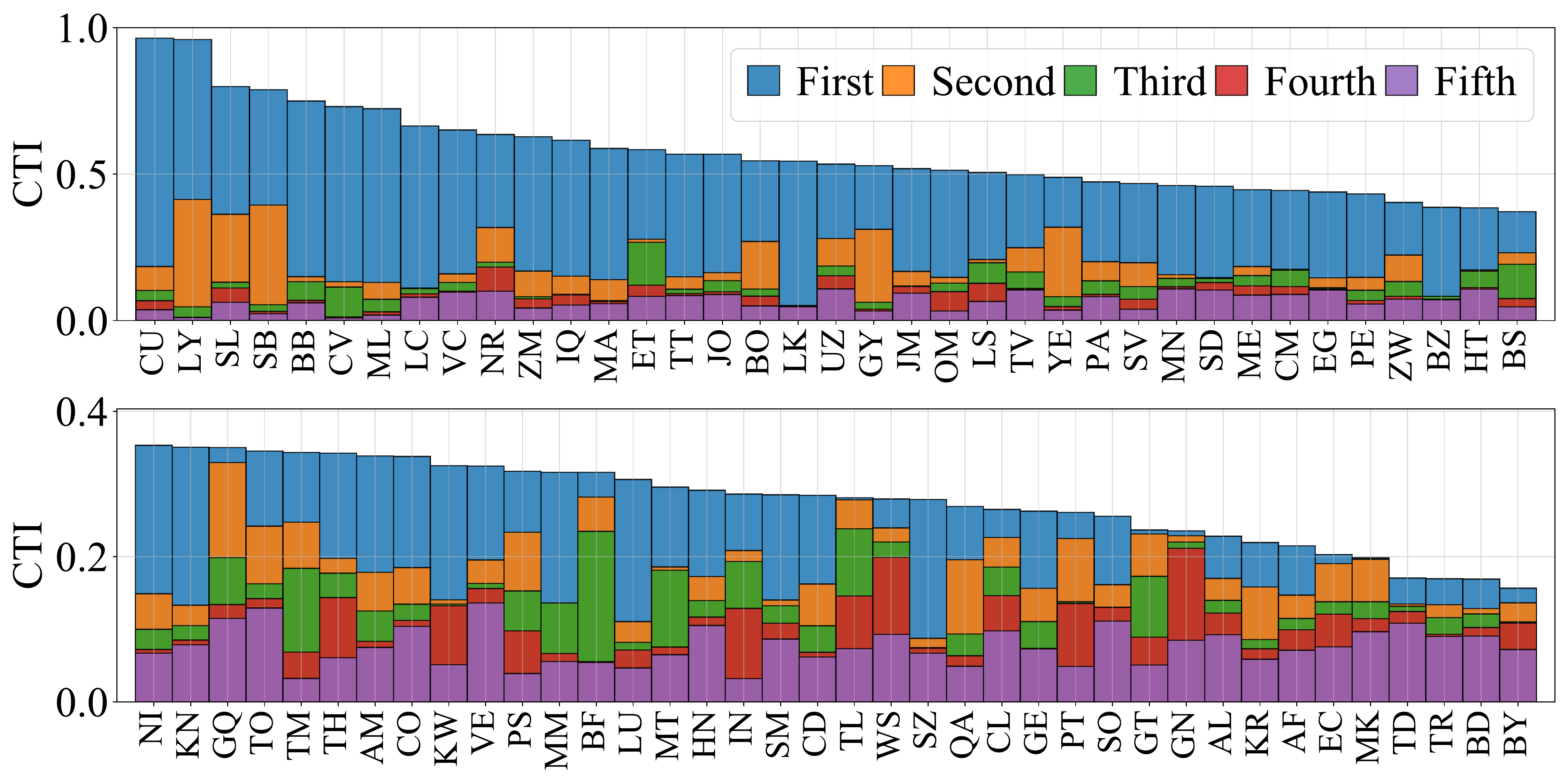}\\
\caption{Overlapping bars showing CTI values of the five top-ranked ASes in the 75 countries we study.
}%
\label{fig:cti-distro}
\end{figure}

Results for the full set of countries we study\footnote{Note that
multiple ASes may provide transit connectivity to the same
prefixes, explaining why the sum of CTI values of top ASes may be greater than 1.} are included in
Fig.~\ref{fig:cti-distro}. We discuss
several representative cases below.

\textbf{Most exposed countries}.
Only four countries have a top-ranked AS with a CTI over
0.75: Cuba, Libya, Sierra Leone, and the Solomon Islands (a small island nation).
Cuba appears to have the most-exposed transit ecosystem\footnote{This 
is consistent with previous work that focused 
exclusively on Cuba, finding its international connectivity to be constrained \cite{Bischof}.}%
, in which
the top-ranked AS has CTI of 0.96. 
Because CTI discounts indirect transit---and the top AS monopolizes
observed, direct connectivity---the CTI of Cuba's remaining ASes declines
rapidly (81\% from the top-ranked AS to the second). 

\textbf{Countries around the median}.
The median of the leftmost bar in Fig. \ref{fig:ctioverview}
consists of countries that are still considerably exposed
to observation and tampering, with CTI values ranging
from 0.34 to 0.44, including: Egypt, Equatorial Guinea, Belize and Thailand.
In Eq. Guinea, %
the top-two ASes each have a CTI over 0.3;
these ASes have a p2c relationship with each other. 
Egypt and Belize
have more skewed distributions, with a 67--79\% decline from the 
top AS to its successor. 

\textbf{Least exposed countries}.
At the other end of the spectrum in Fig.~\ref{fig:cti-distro}
are five countries where the top-ranked has CTI values under 0.2: 
Chad, Bangladesh, Belarus, Turkey and North Macedonia.
These countries have flatter distributions, with CTI declining
at most 21\% (or 16\% on average) between the top-two ASes.
As a result, we find no evidence of these nations being
particularly exposed to a single network (unlike most of their peer countries
in our sample).
India, the country with the most Internet users in our sample, 
is in the bottom third %
with a top-AS CTI of 0.29, declining by 27\% between the top-2 ASes. 

\textbf{Frequently top-ranked ASes}.
Of the 165 ASes present in Fig.~\ref{fig:ctioverview},
126 of them are in the top-5 for only one country, with
a further 31 ASes in the top-5 of at most 10 countries. There
are eight notable exceptions, however:
3356*-Lumen\footnote{Formerly Level3/CenturyLink.}  
(top-5 in 25 countries), 
1299*-Telia (24), 174*-Cogent (24),
6939-HE (18),
5511*-Orange (16),
6762*-T. Italia (14), 23520-C\&W (14), and 6453*-Tata (12).  Nearly
all of these networks (marked with *) are in the inferred clique at
the top of the global transit hierarchy~\cite{ASRankAb8:online}.  C\&W
is only present in our analysis for countries in the Caribbean.  HE
has a very broad footprint, with countries in Africa (7), the
Mid. East (3), W. Europe (2), Southeast Asia (2), South Pacific (2)
and East/South Asia (1 each).

\subsection{Submarine cable operators}
\label{sec:results:cable}

Submarine cables are known to be an important part of the global
Internet
infrastructure~\cite{10.1145/3286062.3286074,Fanou2020,LiuShucheng}
and play a role in the top-5 ASes of most countries we study.
(Nicaragua, Guatemala, and Guyana are the only three nations where none of the
top-5 ASes are associated with the submarine cables landing in the country.)

In this section, for each country, we find the highest-ranked AS by
CTI where there is evidence of an institutional connection between the
AS and an owner or operator of a submarine cable.  We define an AS as
a submarine cable operator if we find a direct match between the AS
Name, the AS Organization \cite{as2org}, or a corporate parent
organization (\eg CenturyLink for Level3, the Government of Sierra
Leone for Sierra Leone Cable Company) and the owners of a submarine
cable operator according to TeleGeography~\cite{telegeography} and
Infrapedia~\cite{infrapedia}.
This process yields submarine cable ASes in 46
countries out of 51 possible, as 17 of the 75 countries are
landlocked, and 7 have no submarine cable connectivity according to
the operator databases.  In three additional countries (Myanmar \cite{myanmar}, 
the Solomon Islands~\cite{solomon},
and Congo DRC~\cite{congodrc}) only TeleGeography provides an AS to
submarine cable match, which we confirm with information from the
cited sources (the operators themselves, the government of Australia,
and a submarine cable news source).
In the remaining two countries (Thailand~\cite{tot} and
Samoa~\cite{samoa}) where we were not able to find an AS to submarine
cable from TeleGeography, we rely on the cited sources (from the
operator and a Samoan news outlet) to find a match.  Note that only
operators of submarine cables who appear as an AS on the BGP path can
be identified using this method. %

\begin{figure}
    \centering
    \includegraphics[width=.92\textwidth]{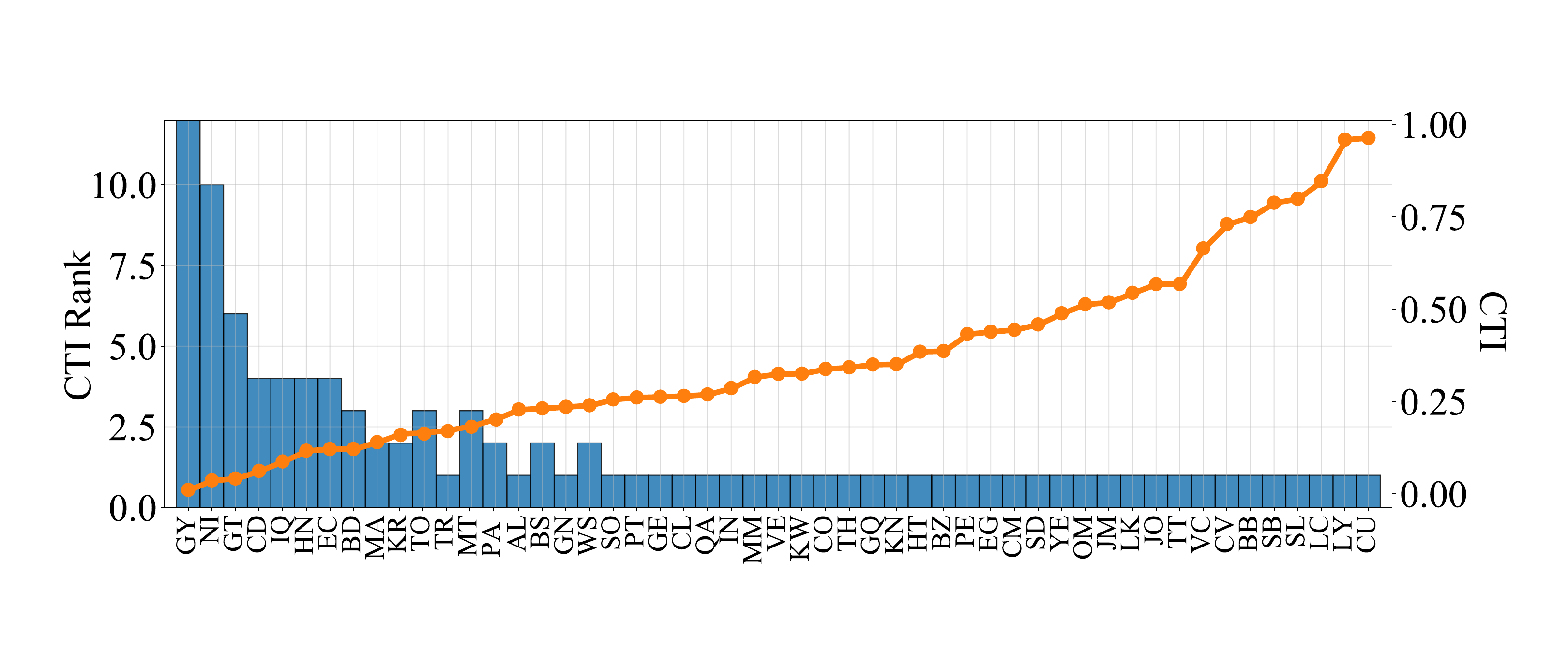}\\
    \vspace{-9pt}
    \caption{Orange circles: CTI of top-ranked submarine cable AS. Blue bars: CTI rank of top-ranked submarine cable AS.}\label{fig:subcable}
\end{figure}

Our findings are shown in Fig.~\ref{fig:subcable}, with the CTI of the
top cable-owning AS in each of the 51 countries shown as orange circles, 
and the ordinal ranking of that AS in its country's ecosystem
as blue bars. %
In 36 countries, a submarine cable AS is ranked at the top by
CTI, with an average rank of 1.9. %

Note that being the top operator by CTI means different things in
different countries, as the underlying potential exposure to observation
affects the CTI of the top AS. For instance, in Turkey %
a cable-owning AS ranks first by CTI, but has the lowest CTI among
such countries. Said AS (9121-Turk Telecom)
has a CTI of 0.17. 
By contrast, in
Cuba and Libya, a submarine cable operator (11960-ETECSA and
37558-LIT) is also ranked first but with CTIs of 0.96 in both cases. 
As a result,
Turkey
is much less exposed to a single AS
than Cuba and Libya.

We also find regional clusters of high transit influence for the same AS
operating a submarine cable, including C\&W (formerly Columbus Networks), which
is among the top providers in 11 countries in
Central America and the Caribbean thanks to its ownership of the ECFS, 
ARCOS-1 and Fibralink cables. 
Telecom Italia Sparkle, Telefonica and Bharti 
Airtel also have an important transit
presence in the Mediterranean, Latin America, and South Asia
respectively. 
We release a 
complete list of submarine cables 
linked to an AS with high CTI on the paper's repository. \color{black}%

\subsection{State-owned transit providers}
\label{sec:stateowned}

In more than a third (26) of nations, we find that at least one of the top-5
ASes is state-owned, %
motivating us to further examine the total influence of a country's
government on its Internet connectivity.  In particular,
we adapt CTI
to quantify the influence of state-owned conglomerates---as some nations have more than one state-owned AS---and apply it to the 75 countries in our sample.
We use as input a list of ASes that are majority-owned by sovereign states~\cite{carisimoimc21}. The list
was manually verified and encompasses both access and transit ASes.
The dataset includes major telecommunication providers
as well as its sibling networks and subsidiaries.
Using this list, we find 100 state-owned ASes
who operate domestically (\ie where the
state owner and the country of operation are the same) in 41 countries.

\begin{figure}
    \centering
    \includegraphics[width=.8\textwidth]{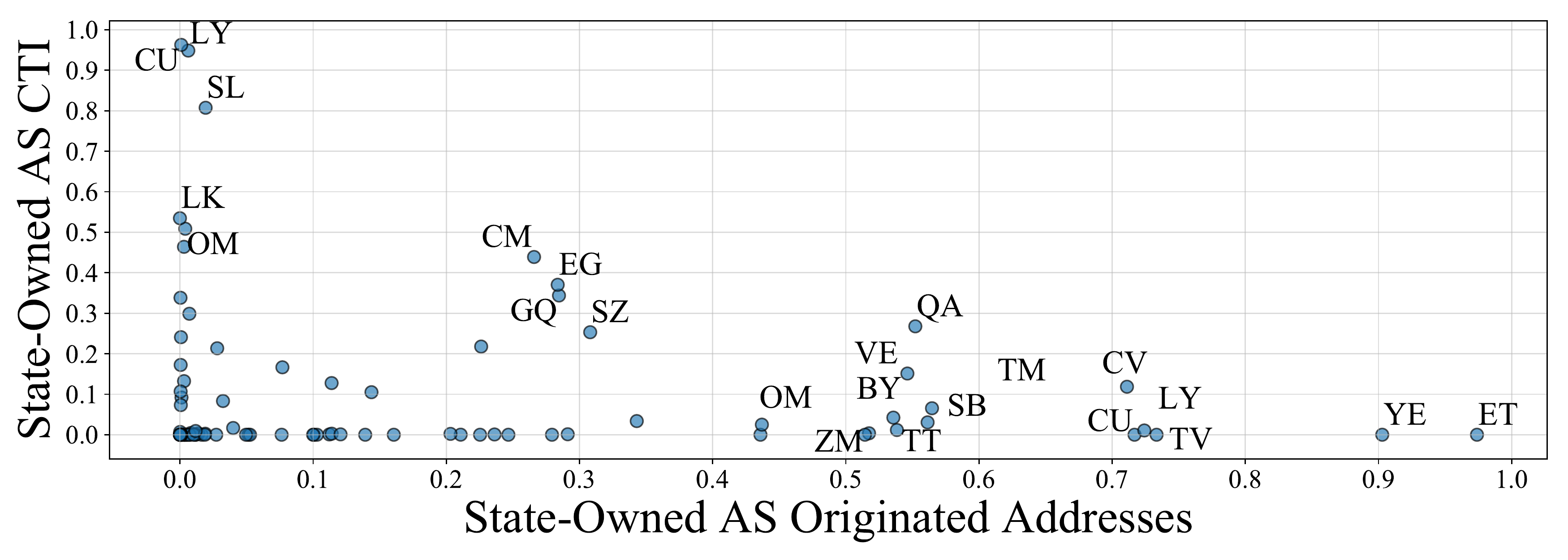}\\
    \caption{CTI and fraction of addresses originated by domestic, state-owned ASes in our study.}\label{fig:statescatter}
\end{figure}

\subsubsection{Influence of state-owned ASes.}    
Our initial exploration of the influence of state-owned ASes concerns
the role each AS plays in the ecosystem of its country, as shown in
Fig.~\ref{fig:statescatter}.  We find that state-owned ASes tend to
provide either transit or access, usually not a combination of
both. 
(Most points in Fig.~\ref{fig:statescatter} line up along an axis,
rather than towards the middle.) 
As a consequence, meaningfully
estimating the footprint of the state requires combining the two kinds of
influence as well as aggregating data for AS conglomerates.
(Two exceptions where a state-owned AS 
provides both Internet access (\ie as an origin AS) and serves
transit to other ASes are Cameroon and Egypt; in the former, Camtel
has both a high CTI (0.44, ranked first) and originates 27\% of the
country's addresses (second only to Orange Cameroon).
Egypt's TE has a CTI of 0.37 and originates 28\% of the country's addresses.)

We begin our combined estimation by computing CTI for not just a single AS, but a
set of ASes, while not ``double counting'' influence over the same
addresses; \ie if two of the state's ASes originate and provide transit to
the same addresses, we add those addresses to the state's footprint
once. We call this derived metric $CTIn$.  Intuitively,
$CTIn$ reflects the ``pure-transit'' footprint of the state, crediting
only the addresses where state-owned ASes serve exclusively as transit
providers.  For instance, if AS $A$ and AS $B$ (both of which operate in
country $C$) respectively originate and provide transit to the same /24
prefix, $CTIn$ says that the conglomerate $S_C = \{A , B\}$
does not have transit influence over the /24 prefix.
Formally, $CTIn_M(S_c, C) \in [0,1]$ is calculated as
\begin{equation*}
\sum_{m\in M} \left( \frac{w(m)}{|M|} \cdot  \sum_{{p} | \mathrm{onpath}^*(S_c,m,p)} \left( \dfrac{a(p,C)}{A(C)} \cdot \dfrac{1}{d^*(S_c,m,p)} \right) \right),
\end{equation*}
\noindent which is essentially identical to Eq.~\ref{eq:cti}, except
that $S_c$ is a set containing all of the ASes in the state-owned
conglomerate of country $C$; $\mathrm{onpath}^*(S_c, m, p)$ is true
if $\mathrm{onpath}(AS_t, m, p)$ is true for some $AS_t \in S_c$ and
$p$ is \textit{not} originated by any AS in $S_c$; and $d^*(S_c, m, p)
= \min_{AS_t\in S_c} d(AS_t, m, p)$, \ie the AS-level distance from
$p$ to the closest AS in the conglomerate.

Finally, we define the total footprint of the state, {\it i.e.}, addresses
that are either originated or for which transit is served
by a state-owned AS.  The state's footprint $F(C) \in [0,1]$ is
calculated as
\begin{equation*}
F(C) = CTIn_M(S_c,C) + \sum_{AS_o \in S_c}  \dfrac{a^*(AS_o,C)}{A(C)},
\end{equation*}
\noindent where $a^*(AS_o,C)/A(C)$ is the fraction of addresses in country
$C$ originated by $AS_o$.  The first term of the sum is the
pure-transit footprint and the second term is the addresses directly
originated by the state-owned conglomerate $S_c$.

\begin{figure}
\vspace{-5mm}
\centering
\includegraphics[width=.8\textwidth]{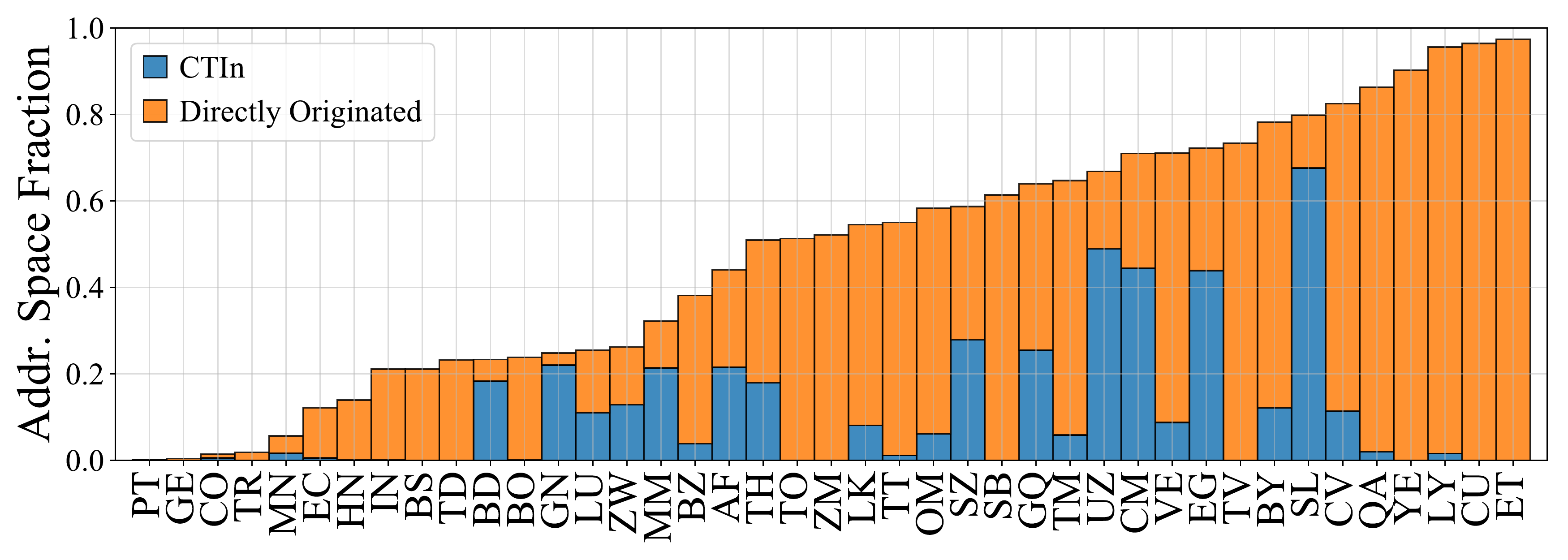}
\caption{State-owned originated address space $a^*$ (orange bars), $CTIn$ (blue bars), and state footprint $F$ (bar height) for countries in our study.
}\label{fig:stateline}
\vspace{-8mm}
\end{figure}

\subsubsection{Findings.} Fig. \ref{fig:stateline} shows our findings for the state-owned
footprint ($F$, bar height), the originated fraction by state-owned ASes
(orange bar), and pure-transit footprint of
state-owned ASes ($CTIn$, blue bar).
Our results suggest that domestic state influence exists on a spectrum
where some countries, such as Ethiopia, Cuba, Libya and Yemen, rely
overwhelmingly on the state for the provision of Internet access and 
($F$ between 0.90--0.97), whereas
others, such as Colombia, Turkey, Mongolia and Ecuador have relatively
marginal state-owned enterprises ($F$ between 0.01--0.12).

Regarding the mode of influence that states use, in many
countries in Fig.~\ref{fig:stateline}, %
most of the bar height is contributed by the orange portion,
meaning that the footprint of the state
comes from addresses directly originated. %
However, in some countries the state punches above its access network
weight by deploying an influential transit provider, \ie those where
the bar height is not dominated by the origin contribution in
orange.

\begin{table}[]
\centering
\caption{Top countries by $CTIn$.\label{tab:ctinonstate}}
\small
\begin{tabular}{p{2.4cm}p{0.9cm}p{0.9cm}p{0.9cm}p{0.9cm}p{0.9cm}p{0.9cm}p{0.9cm}p{0.9cm}p{0.9cm}}
\toprule
Country & SL & UZ & CM & EG & SZ & GQ & GN & AF & MM\\
\hline
$CTIn$ & 0.68 & 0.49 & 0.44 & 0.44 & 0.28 & 0.26 & 0.22 & 0.21 & 0.21\\
\hline
$F$ & 0.80 & 0.67 & 0.71 & 0.72 & 0.59 & 0.64 & 0.25 & 0.44 & 0.32\\
\hline
\end{tabular}
\end{table}

\subsubsection{Pure-transit footprint of state-owned ASes.}
The countries where pure-transit influence ($CTIn$) 
is largest (0.2 or more, or pure-transit influence over
at least a fifth of the country's addresses) 
are shown in Tab. \ref{tab:ctinonstate}.
In these countries, all of which are in Africa and Central Asia,
providing transit considerably increases the influence of the state.
We note that the mere existence of these influential transit ASes
does not signal willingness of the state to engage in surveillance or selective
tampering, but rather that the government may have opportunities 
to do so. For instance, Myanmar's state-owned 
Myanma Posts and Telecommunications (MPT),
which is included in our analysis,
appears to have been involved in the disruption
of the country's Internet service during the recent coup \cite{Myanmarc48:online}.

\section{Inferring Transit Dominance}
\label{sec:transit}

In this section, we describe how we identified the 75 countries that are the focus of the preceding section, \ie countries where 
provider-customer transit (p2c) relationships
are likely the dominant mode of inbound
international connectivity.
We start by identifying countries for which public datasets of
Internet Exchange Points (IXPs) and Private Colocation facilities
(Colo) show no evidence of international peering
(Sec.~\ref{sec:peeringtest}).
Based on this analysis, we
conduct an active measurement campaign to confirm the absence of
international peering (Sec.~\ref{sec:campaign}).
This second stage based on traceroutes is necessary because peering
datasets are incomplete, particularly when it comes to membership
lists at IXPs in developing countries~\cite{lodhi}.  We consider
the prevalence of transit links being used to reach each of our target
countries from probes distributed worldwide (\S~\ref{sec:finalcountries})
in combination with our operator validation (\S\ref{sec:validationmain})
\color{black} 
to select a set of transit-dominant countries. %

We define international peering 
as a (logical) link between two ASes that: \ione operate primarily in different
countries (Sec. \ref{sec:nationality}), and \itwo where that link is not an
inferred transit-customer link. We use this definition since we are interested
in studying the AS-level routes taken towards each country.
We are aware of the limitations of our measurements and analysis, particularly
with regards to the location (both topologically and geographically) of our
probes; we address the issue further in Sec. \ref{sec:limitations}.

\subsection{Constructing a candidate list}
\label{sec:peeringtest}

\label{sec:nonpeering}
We identify countries where international peering may not be
prevalent
by evaluating evidence of international peering involving origin ASes
present in the country. While domestic peering is very common, our
hypothesis is that international peering is still not a frequent
occurrence in some countries.
We begin with the set of ASes that originate at least 0.05\% of
addresses in each country. We remove marginal ASes that originate a
very small fraction of the country's address space to reduce the scope
of our active campaign, as we are limited by RIPE Atlas's system-wide
limits on concurrent measurements~\cite{ripeudm}.  This set includes
origin ASes that we classified as foreign to that country, but that
originate BGP prefixes entirely geolocated in the country. (These ASes
originate a marginal fraction of the addresses in the vast majority
of countries we study; see \S\ref{sec:geolocation}).  We look
for these origin ASes in CAIDA's IXP dataset
(from Oct. 2019~\cite{caidaixp}), PeeringDB Colo dataset (from Mar. 1st,
2020~\cite{peeringdbpni}), and inferred AS-Relationships from BGP
(Mar. 2020~\cite{asrelationships}).

We classify an origin AS as a \textit{candidate} if the following
three conditions are true:
\renewcommand{\labelenumii}{\Roman{enumi}}
\begin{enumerate}
  \item the origin AS has no foreign peers in BGP~\cite{asrelationships};
  \item the origin AS is not a member of any IXPs or Colos based in another country~\cite{caidaixp,peeringdbpni}; and
  \item the origin AS is not a member of any IXPs or Colos
  where any member AS is based in a different country than the origin AS~\cite{caidaixp,peeringdbpni}.
\end{enumerate}

\noindent The intuition for each test is as follows. If we observe at least one foreign
peer on BGP (1), this origin AS already has the ability to receive some external
content from that peer, bypassing transit providers. Therefore,
transit providers serving that origin will have fewer capabilities to 
observe traffic flowing towards it. %
Further, if an AS is a member of an IXP/Colo in another country (2), or a member
of an IXP/Colo where another member is from a different country (3), the
origin AS is at least capable of establishing peering
relationships with those other ASes.  

\begin{figure*}[htpb] \centering
\vspace{-4mm}
\subcaptionbox{\label{fig:peeringheat}}{%
    \includegraphics[width=.32\textwidth]{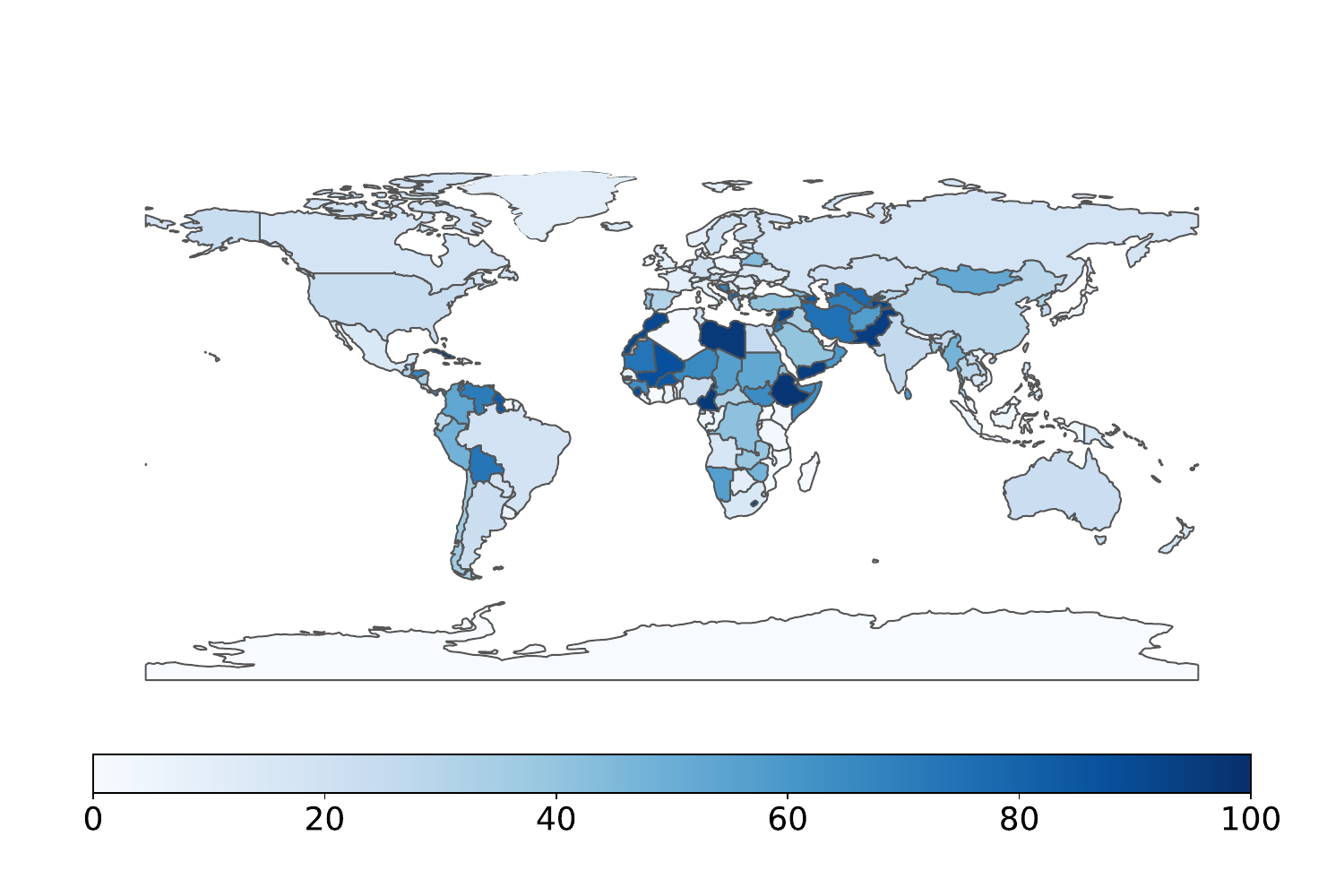}
}~\subcaptionbox{\label{fig:transitcountryfrac}}{%
    \includegraphics[width=.32\textwidth]{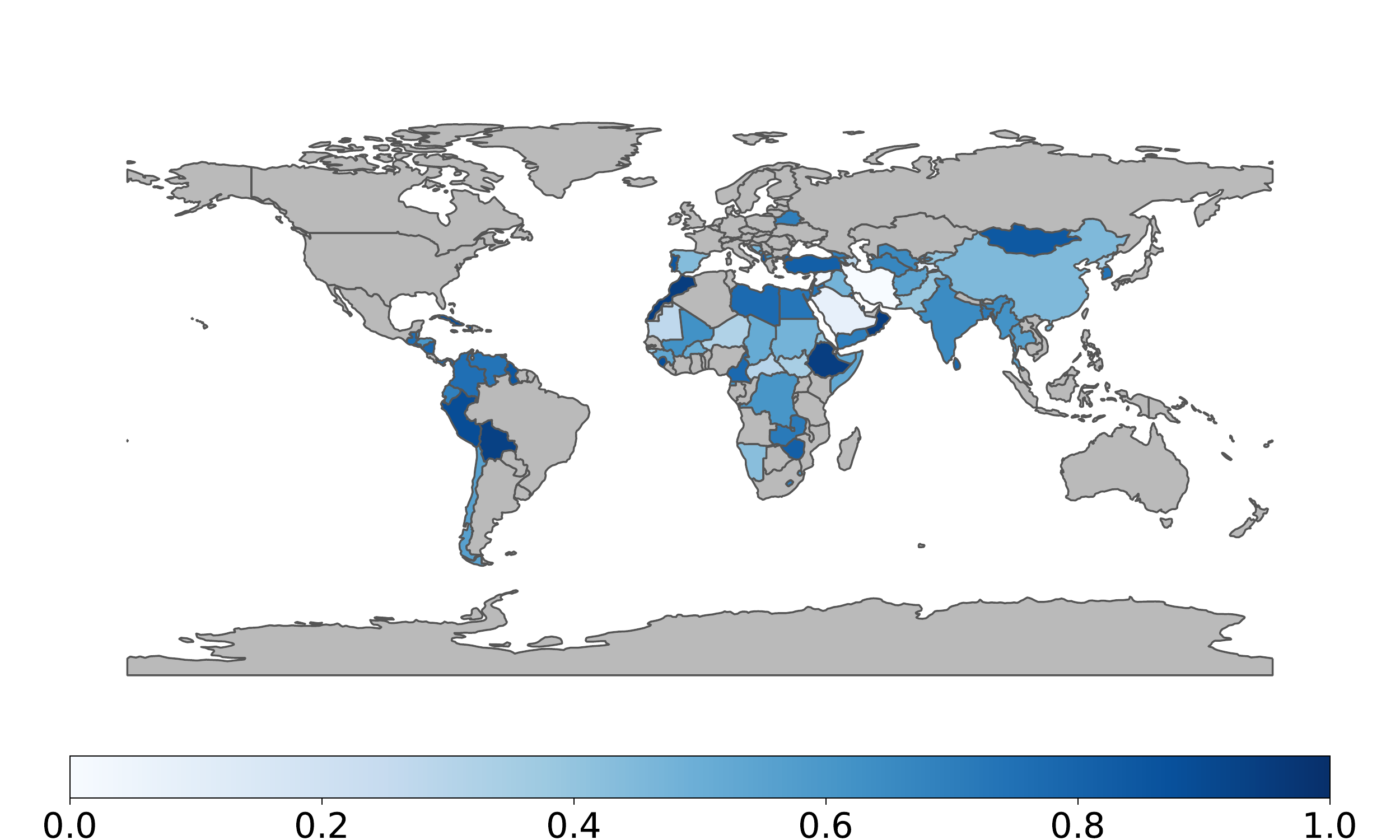}
}~\subcaptionbox{\label{fig:activetest}}{%
    \includegraphics[width=.32\textwidth]{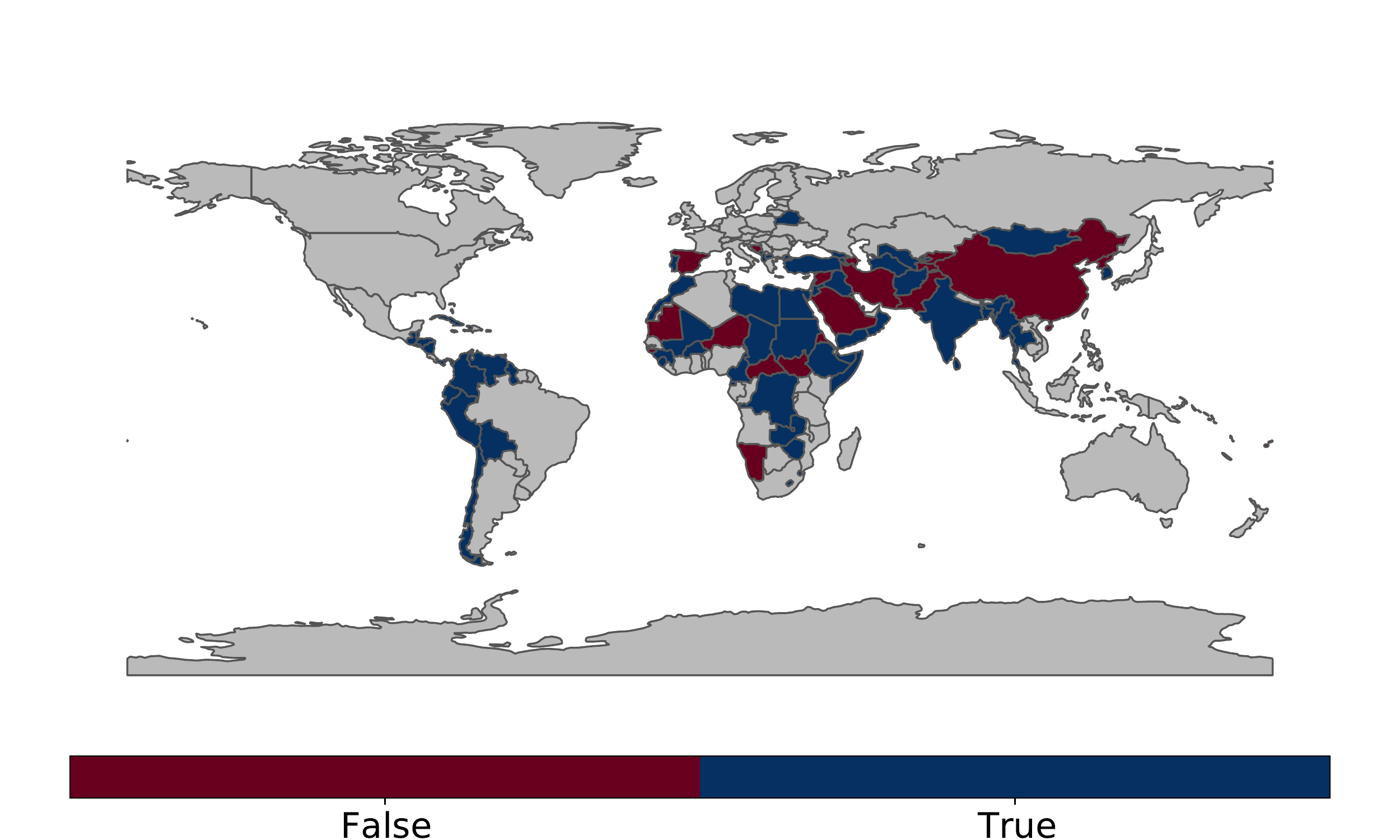}
}
\caption{Non-peering observed perc. on passive datasets \ref{fig:peeringheat}, scaled country-level transit fraction in probed countries \ref{fig:transitcountryfrac}, and  final set, with countries in red excluded \ref{fig:activetest}.}
\label{fig:heatmaps}
\vspace{-8mm}
\end{figure*}

Fig.~\ref{fig:peeringheat} shows the percentage of a country's address
space originated by candidate ASes.  We select the top-100 countries
as candidates for active measurements. This set includes only
countries where at least 25\% of addresses are originated by candidate
ASes. Our motivation is to actively probe the set of countries where
it is most likely that transit providers still play an important role
on inbound international connectivity.  These 100 countries are
colored in Fig.~\ref{fig:transitcountryfrac}.

\subsection{Active measurement campaign}
\label{sec:campaign}

We ran a traceroute campaign to the 100 candidate countries for 14
days starting May 2nd, 2020. %
Additionally, we use all publicly available IPv4 traceroutes on RIPE Atlas
during the same period---on
the order of several million per hour---in order to opportunistically
take advantage of other measurements towards the same ASes.
\label{sec:traceroute}
We design our traceroute campaign guided by two constraints. First, we
want to select a geographically and topologically diverse set of
probes. %
Second, we have to operate within the rate limits of RIPE
Atlas\footnote{Which RIPE Atlas generously relaxed for this study upon
direct request.}, particularly regarding concurrent measurements
and credit expenditure.

Within these constraints, we launch ICMP traceroutes\footnote{Using
default RIPE Atlas values except number of packets (reduced to 1).}
from 100 active---shown as ``connected'' during the previous
day \cite{ripestatus}---RIPE Atlas probes (located outside any target country)
towards a single destination in each AS, twice daily\footnote{We
space traceroutes an hour apart in 800-target IP blocks.}; %
probing at this frequency gives us
28 opportunities to reach the AS during the two-week period from each vantage
point.

We target an IP in a single /24 block for each origin AS in each candidate
country by looking for any prefix originated by that AS that is entirely
geolocated or delegated within the candidate country (see Sec.
\ref{sec:geolocation}).  Our final dataset is comprised of 33,045,982 traceroutes,
including those launched by other RIPE users that meet our constraints. 
The distribution of the number of traceroutes reaching each country 
has the following properties: 
(Min, 25th Pctl., Median, Mean, 75th Pctl., Max) = (36, 13k, 46k, 330k,
250k, 3.3m). That is, the median country received 46k traceroutes. 
Only three countries received fewer than a thousand traceroutes: 
Eritrea (667), Nauru (154), and Tuvalu (36).

\label{sec:aslevel}

We use BdrmapIT~\cite{bdrmapit} to
translate our traceroutes into AS-level interconnections.
BdrmapIT requires a number of external datasets in its operation, which
we specify as follows: inferred AS-Level customer
cone~\cite{Luckie2013} from Mar. 2020; \textit{AS2Org}, which infers
groups of ASes who belong to the same organization\footnote{This
dataset is published quarterly.}, from Jan. 2020; and datasets we
mention in other sections---prefix-to-Autonomous System
mappings (\S\ref{sec:method}), \textit{PeeringDB} records
(\S\ref{sec:peeringtest}), and RIR delegation records
(\S\ref{sec:geolocation}).
From these traceroutes and external datasets,
\textit{BdrmapIT} infers a set of
AS-level interconnections and the IP addresses (interfaces) at which
they occur.  Each interface inferred by \textit{BdrmapIT} has an AS
``owner'' assignment. We reconstruct the AS-level path observed on the
traceroute using such assignments.

\subsection{Country-level transit fraction}
\label{sec:finalcountries}
From the preceding sections we have built a set of AS-level paths
taken from the traceroute source to the destination AS.
We now need a quantitative analysis technique to infer the prevalence of
transit links on inbound traces towards each country.

To that end, we determine how frequently a transit (p2c) link is traversed when
crossing the AS-level national boundary\footnote{As defined by our AS
Nationality (\S\ref{sec:geolocation}), not actual political borders.} towards an origin AS
($AS_o$) in a candidate country.  We infer the AS-level national
boundary as the link between the last foreign AS observed on the
AS-level path (starting from the vantage point) and the subsequent AS.

We calculate how frequently, in the inbound traceroutes we process with
\textit{BdrmapIT}, the AS-level national border crossing occurs on a transit link
for each origin AS.
We scale this fraction to take into account the size of the address space
originated by each AS using the \textit{country-level transit fraction}:
\begin{equation*}
\label{eq:countrytransit}
T(C) =\sum_{{AS_o,AS_c}\in \mathrm{dom}(C)} \sum_{AS_t \notin \mathrm{dom(C)}} \dfrac{R(AS_o,AS_t,AS_c)}{R(AS_o)} \cdot \dfrac{a^*(AS_o,C)}{A(C)},
\end{equation*}
\noindent where $R(AS_o,AS_t,AS_c)$ is the number of traceroutes
destined toward a prefix originated by $AS_o$ that traverse a transit
link between a foreign provider $AS_t$ and a domestic customer $AS_c$
in country $C$;
$R(AS_o)$ is the total number of traceroutes where $AS_o$ is the last
observed AS; and $a^*(AS_o,C)/A(C)$ is the fraction of country $C$'s
address space originated by $AS_o$.  %
For instance, if an AS originates 50\% of the country's origin
addresses, and 50\% of the traces towards it traverse a foreign transit provider
AS, the contribution of that AS to the country-level transit fraction becomes
0.25.
Note that $AS_c$ and $AS_o$ are not necessarily the same,
as the border crossing may occur at the link between (direct and/or indirect)
providers of $AS_o$.

The values of $T(C)$ for each candidate country are represented in Fig.
\ref{fig:transitcountryfrac}: countries in darker shades of blue have both a
large probed and responsive fraction and a large fraction of traceroutes from
outside the country traversing transit providers.  
The closer the fraction is to 1, the more evidence we have that the
country relies on transit providers for its international inbound connectivity.

\begin{figure}
    \centering
    \includegraphics[width=.8\textwidth]{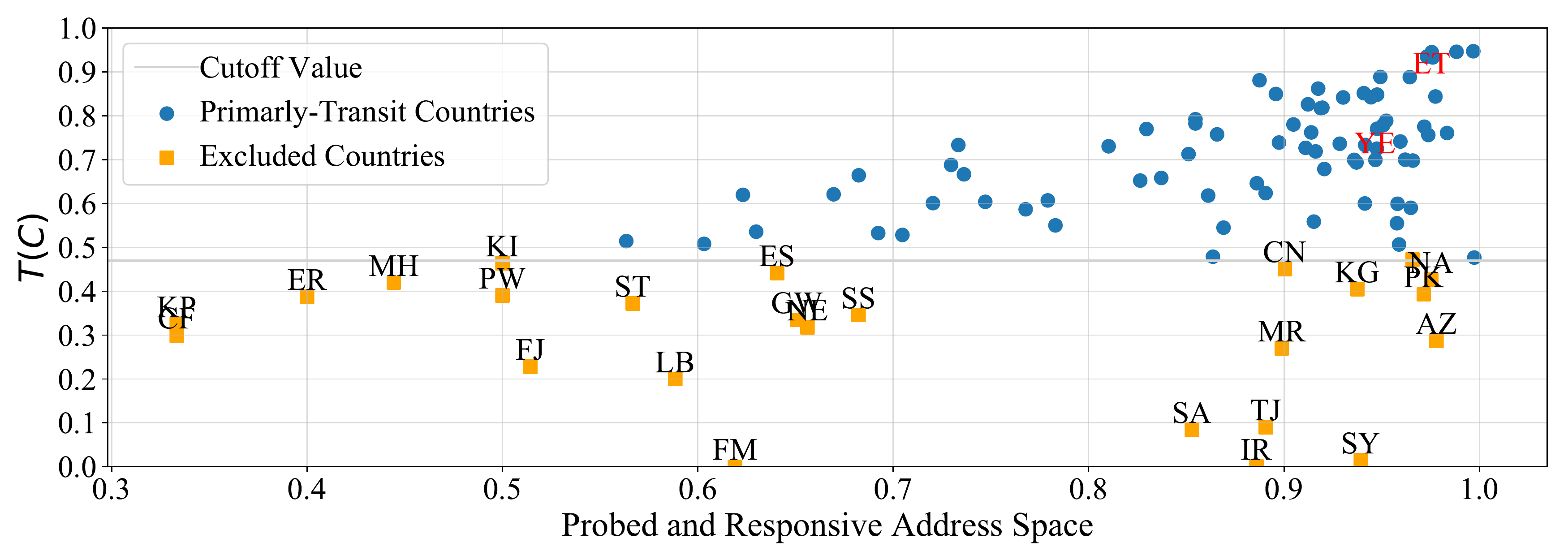}\\
    \caption{Country-level transit fractions $T(C)$ for countries in our sample.}\label{fig:ctitestline}
\end{figure}

\subsection{Final selection}
Finally, in order to identify a set of primarily-transit countries, 
we evaluate the values of $T(C)$
across countries, shown in Fig. \ref{fig:ctitestline}.
At one extreme of Fig. \ref{fig:ctitestline} and Fig. \ref{fig:transitcountryfrac}
are countries such as Ethiopia (ET) and Yemen (YE), $T(C)$ = 0.95 and 0.7,
respectively, where all available evidence points towards transit links as the
main inbound modality.  At the other extreme are countries such as Syria (SY) and
Iran (IR), $T(C) \leq$ 0.01, where we rarely observe AS-level
national borders being crossed using transit links.

Outside the upper and lower extremes in Fig. \ref{fig:ctitestline}, where the
decision of whether to include a country in our study is obvious, the middle
results (most countries) do not offer clear dividing points.  We decided then to
set the threshold for $T(C)$ to classify a country as primarily-transit based on our
validation with operators (\S\ref{sec:validationmain}); in particular, we use the
value of $T(C)$ for Sudan (0.48) as a lower bound, which is the
lowest $T(C)$ in any country that we were able to confirm relies on transit
links for its inbound connectivity.
The final countries in our CTI study are shown in a blue-white spectrum in Fig.
\ref{fig:activetest} and as blue circles in Fig. \ref{fig:ctitestline}, 
75 of the 100 
candidates. %
Countries in red are excluded from further analysis,
as at this time we lack sufficient evidence to 
support that they are primarily using transit providers for inbound
connectivity.

\vspace{-0.2cm}

\section{Stability and Validation}
\label{sec:validationmain}
In this section, we discuss the findings of 
our stability analyses, validation with operators,
and a calculation of transit influence at the organization level.

\subsection{Stability}
\textbf{Temporal stability}. 
We apply our CTI methodology to a set of BGP paths from Feb. 2020 and Apr. 2020 and
compare the results to those presented in \S\ref{sec:all-results} (from Mar. 2020).
Specifically, we compute the absolute value of the difference in CTI
across successive months for transit ASes listed in each country. 
The results are shown in Tab. \ref{tab:validation-summary}.
We find that the CTI values are
relatively stable across these months.
\color{black}

\begin{table}[]
\vspace{-5mm}
\caption{CTI Temporal Stability Analyses.}
\small
    \centering
    \begin{tabular}{r|c|c|c|c|c}
	\hline
 	\textbf{Type} & \textbf{Compared Sets} & \textbf{25th ptile.} & \textbf{Mean} & \textbf{Median} & \textbf{75th ptile} \\
	\hline
	\textbf{Temporal} & Feb. \& Mar. 2020& 0.00000 & 0.00190 & 0.00001 & 0.00016 \\
	\textbf{All ASes} & Mar. \& Apr. 2020& 0.00000 & 0.00156 & 0.00001 & 0.00017 \\
	\hline
    \end{tabular}
   \label{tab:validation-summary}
\vspace{-5mm}
\end{table}

\textbf{Stability to changes in geolocation input.}
In order to assess the potential fragility of our study to inaccuracies in geolocation, we also applied our CTI methodology using MaxMind~\cite{maxmindgeoloc} and computed the absolute value of the difference in CTI
scores produced with each location database.
The output of
this analysis is (25th perc.,mean,median,75th perc.) =
0.00000, 0.00104, 0.00002, 0.00017, suggesting CTI is
relatively stable across these geolocation inputs.
\color{black}

\subsection{Operator Validation}

We discussed our findings with employees or contractors
of two types of organizations: commercial network operators
and non-profits who conduct networking research (universities,
registrars, and non-commercial network operators).
Additionally, we describe the results of our discussions follwing a mass email
request to ASes with prefixes geolocated in countries in our study.
Discussions with all of these organizations are anonymized.
Our findings are largely consistent with each operator's view of the transit ecosystem
of the countries discussed with them.

The results of our discussion of CTI findings with 6 operators
in 6 countries\footnote{We sent a set of ASes produced before 
updating our CTI methodology to its current form, which explains the ``unconfirmed'' column;
the ``top'' ASes were defined as the country's top 12, unless any of those ASes had a marginal CTI score.}
are shown in Tab. \ref{tab:validation-operator}.
\color{black}
Our CTI operator discussions consist of a
confirmation of the AS set we identified as being
most influential in their countries.
\color{black}
Overall, operators confirm that the vast majority of ASes we identify are 
among the most influential in their nations.
We also summarize our discussions with: \ione operators regarding
our inferences of transit-dominant countries,
\itwo ASes with prefixes geolocated to these countries.
Regarding \ione 10 operators in 9 countries\footnote{CO, ET, CD, LS, SZ, ZW, VE, SD and CM.}
confirmed that their nations are primarily transit\footnote{
\color{black}
Sample, anonymized operator response:
``Sudan is characterized by the traditional IP transit model.
There is a domestic IXP, which serves five ISPs and [redacted AS Name]'s DNS nodes, but there are no foreign network operators present here.
Furthermore, until recently, only two ISPs held gateway licenses (\ie were licensed to provide external connectivity to Sudan).''}.

Regarding \itwo we sent a mass email request to the WHOIS \texttt{abuse} address registered by
ASes that had prefixes geolocated in
10 countries\footnote{We only contacted ASes
who had $\geq1$\% of their addresses in the country. Since
this survey took place in 2021, we use the addresses geolocated in Jan. of that year.}
(with IRB approval):
BO, CO, VE, CM, BD, GT, CL, HN, SV and ZW\footnote{Selected as a mix of large \& small (by \#ASes) EN- and ES-speaking countries.}.
We received 111 responses in 9 of these countries (all but ZW).
Of these, 107 confirmed they operate primarily in the country
that we geolocated their prefixes to\footnote{In 3 cases, they
stated that they operate in multiple countries.}.
Additionally, 108 were willing to discuss which
type of business relationship dominated their inbound
international traffic: 83 stated that
transit relationships are the primary modality.

\begin{table}[]
\centering
\vspace{-5mm}
\caption{CTI operator validation in 6 countries: CO, ET, ZW, SD, CD and CM.}
\small
    \begin{tabular}{r|c|c|c|c}
	\hline
	\textbf{AS-Country Pairs} & \textbf{\#Confirmed} & \textbf{\#Rejected} & \textbf{\#Unconfirmed} & \textbf{Total \#ASes} \\
	\hline
	\textbf{Top 5 ASes} & 27 (90\%) & 1 (3\%) & 2 (7\%) & 30   \\
	\textbf{All Top ASes} & 45 (79\%) & 7 (12\%) & 5 (9\%) & 57 \\
	\hline
   \end{tabular}
   \label{tab:validation-operator}
\vspace{-5mm}
\end{table}

\subsection{Organization-Level Transit Influence}
In some instances, multiple ASes may be operated by the same
organization.  We identified 323 instances where multiple ASes belonging
the same organization (as of Jul. 2020 \cite{as2org}) have $CTI > 0$
in a given country.  We compute an upper bound of the organization's
transit influence (in each country) by summing the CTI of component
ASes. We find that 270 org-country pairs---an organization operating in
a country---have marginal influence, with the CTI sum under 0.05 (218
were under 0.01).  

For the remaining 53 organization-country pairs, we
compute the contribution to the CTI sum of the highest-ranked AS in
each organization. We separate these into three groups: \ione In 36
org-country pairs, the top AS contributes at least 90\% of the CTI sum
(98\% on average). %
In these 36 cases,
then, a single AS is responsible for the vast majority of the
organization's transit influence.
\itwo In 7 org-country pairs, the contribution to the CTI sum
of the additional ASes---other than the top AS---in the organization is between 
0.01--0.04 (between 11-29\% of the CTI sum), or 0.02 on average. 
Therefore, %
the change in CTI as a result of their 
inclusion is relatively marginal.

\ithree In the remaining 10 org-country pairs, only 4 have a CTI sum greater than 0.1.
For these, we compute the $CTIn$ of the organization
to determine the contribution of the top AS in each organization (rather than a lower bound).
In all 4 cases, the top AS contributes 61\% or more of the organization's $CTIn$ (country-org, perc. of $CTIn$):
VE-Lumen (87\% of 0.16), SZ-Orange (61\% of 0.14), WS-Lumen (73\% of 0.30), and TV-Internap (62\% of 0.11).
Three of these countries are either a microstate (SZ) or a small island nation (WS and TV). The last instance,
in Venezuela, is likely a consequence of the merger of 
two large companies: AS3356 (Level 3) and AS3549 (Global Crossing) \cite{globalcrossing}.

\color{black}

\section{Limitations}
\label{sec:limitations}

At a high level, CTI assumes all ASes and IP addresses are equivalent, which is certainly not the case.
At the AS level, it is possible that one, dominant AS provides
stronger security than a multitude of smaller ASes with tighter
budgets. From the perspective of an attacker, though, a single AS
having high CTI creates an opportunity; in the case of sophisticated
attackers such as nation-states, the possibility of infiltration of
any network cannot be discarded, but compromising many ASes
simultaneously---in order to observe traffic towards countries where no
AS has high CTI---may be more challenging. As such, ASes
with very high CTI still present a concerningly large observation
footprint, regardless of their level 
of security against infiltration\footnote{\color{black}Recall that CTI 
studies exposure to 
\textit{inbound} traffic observation or selective tampering,
which is unaffected by potentially asymmetric AS paths.}. %

Similarly, IP addresses can represent vastly different entities.  Both
access and transit ASes may deploy carrier-grade network access
translation (CGNAT) \cite{10.1145/2987443.2987474}. Since our model
treats all routed IPs equally, it does not currently take into account
the number of hosts multiplexing a single IP address. We leave this to
future work, but note that an additional weight may be added to CTI:
one that scales up the number of IP addresses in a given prefix by the
number of hosts\color{black}---or the number of ``eyeballs''---\color{black}connected 
to those IPs, on aggregate.  Even within a
given network, however, individual hosts are unlikely to be equally
important as some (e.g., those belonging to governmental organizations
or power-grid operators) may have more sensitive traffic.  Conversely,
some networks might not even actually use all their IP
addresses---although the latter issue is likely less of a concern
in the countries we have studied as their allocation of IPv4 addresses 
tends to be constrained~\cite{Dainotti14}.

In addition to this fundamental conceptual limitation, there are a
variety of technical details that could have out-sized impact on our conclusions:

\emph{Incomplete BGP data.}
We acknowledge that the BGP paths we observe and use to compute CTI
are incomplete given the location of BGP monitors.
Given the serious implications for countries that appear highly exposed to external observation and selective tampering 
by an AS, we argue that it is important to study such exposure with available data.
Further, we note that there are two important factors aiding the credibility of our
CTI findings:
\ione our validation with network operators, who have confirmed that the
set of transit ASes identified in their countries is largely consistent with 
their own understanding of the country's routing ecosystem. 
\itwo There is
greater visibility over p2c links in the AS-level topology \cite{Luckie2013,dhamdhereevo},
which enables our analysis as we are studying exposure to observation 
or selective tampering by transit ASes, in particular.

\color{black}
Despite these mitigating factors, we recognize that BGP incompleteness
may impact the accuracy of CTI findings. We leave to future work
an analysis of CTI's sensitivity to changes in the BGP input
(which would further mitigate concerns with BGP incompleteness),
\eg the addition or removal of BGP monitors, or the 
addition or removal of ASes who feed into each monitor.
Finally, we note that CTI incorporates an outlier filter
(\S \ref{sec:hegemony} and \S \ref{sec:hegcomparesec}) 
which has been shown as robust to changes in BGP input 
monitors~\cite{Hegemony}.

\color{black}

\emph{Traffic.}
We use a country's geolocated IP(v4) addresses as a proxy for
the nation's traffic, as this is a limited resource that is necessary to
connect any device to the Internet.  IP addresses are often used as a proxy for
traffic, \eg in \cite{soldo}, and previous work has found strong correlations
between number of IP addresses observed in BGP and traffic volume for ASes that
provide either access or transit service \cite{lodhi}.  An AS that serves a
larger number of IP addresses would consequently have more capabilities for
traffic observation, either of a larger share of potential
devices, or of traffic that is more sensitive in nature. 

Additionally, we do not study direct peering with cloud/content providers,
who are responsible for large volumes of user-destined traffic.
In addition to p2p links with access or transit networks,
these content providers may have in-network caches in 
the countries we study. These caches may be placed
in the access network itself, 
in the influential transit providers we have identified,
or elsewhere~\cite{10.1145/3211852.3211857}. 
Content providers are large and complex distributed systems,
employing sophisticated load balancing~\cite{10.1007/978-3-319-17172-2_7}, 
routing, and DNS~\cite{10.1145/1159913.1159962} techniques.
Given these complexities, we leave to future work an evaluation
of the impact on CTI of
direct peering with cloud/content providers, 
and in-network cache placements.
\color{black}

\emph{Imperfect geolocation.}
A potential source of inaccuracy is IP geolocation, as assigning prefixes
to a geographic area is challenging and the commercial providers
who sell such information use proprietary methods.
We have mitigated these concerns by calculating CTI using two 
commercial providers (\S\ref{sec:geolocation}), 
and find that the metric remains stable.
We have also limited our analysis to the country level,
where geolocation is more accurate than at finer granularities~\cite{Huffaker,Poese,Gharaibeh}.
\color{black}
Further, while determining the location of prefixes originated by large
transit providers with a global presence is problematic because of 
its dynamic nature and wide geographic spread, most networks 
are much smaller and will have limited geographic presence 
beyond their primary country of operation \cite{ZhuoRan} (where 
most or all of their addresses
will be located).

\emph{IPv6.}
Finally, we note that although our model has so far only been 
applied to IPv4 addresses---a reasonable scope given that IPv6 deployment 
is far from wide in many developing regions, 
including Africa \cite{agbaraji2012ipv6,Livadariu}---the code libraries and software
tools we have used are compatible with IPv6, enabling
future research in this area.  

\emph{Inferring Primarily-Transit Countries.}
Any active campaign launched using publicly available
infrastructure will be limited in its effectiveness to reveal peering links by the 
location of vantage points (VPs) from which the traceroutes are launched.
Our campaign is no exception: our VPs are located in a small subset
of the world's ASes, and primarily in Europe and North America.
However, we argue that our measurements form a sufficient basis
to infer that, in the countries we have identified, foreign
peering is rare, since: \ione we discussed our findings with operators
in 12\% of these countries, all of whom have confirmed that
their nation relies primarily on transit providers to receive traffic
from other countries since foreign peering there is rare to nonexistent;
\itwo while our measurements are launched primarily from the U.S.
and Europe, these regions do serve as important content sources
and transit hubs (incl. for intracontinental traffic) for countries 
in Latin America, the Caribbean and Africa \cite{Bischof,galperin2013connectivity,Kiedanski,WhyMiami14:online,fanou2},
where most of the nations we have identified are located.

\section{Related Work}
\label{sec:related}
Several previous studies have focused on 
country-level routing, both for the identification 
of topological bottlenecks \cite{Roberts,Leyba} 
and to evaluate the impact of specific countries' ASes
on routes towards other countries \cite{Karlin}. All of these 
studies have used delegation data
to map an entire AS to a country;
these inferences are prone to inaccuracies when compared with more accurate and granular 
data such as IP-level geolocation, as important transit ASes
may span multiple or many countries, or operate in a country 
different from their registration. 

Previous work focused on the topologies of specific countries
(Germany \cite{Matthias} and China \cite{Zhou2008}) and
relied on country-specific methods and data sets
that do not generalize to automatic inference of AS 
influence in any given country.
Fanou {\em et al.} \cite{fanoupaths} studied the interdomain connectivity 
of intracontinental paths in Africa, using a large traceroute campaign (rather than BGP paths).

CAIDA's AS Rank~\cite{Luckie2013} is another topological metric
developed to characterize the customer footprint of
an AS on the global routing system. %
It does not try to capture the capabilities for observation
of a transit AS for traffic flowing towards a country; we developed the CTI metric to
try to do so. %

\subsection{National Chokepoint Potential and Hegemony}
\label{sec:hegcomparesec}
\color{black}
In this subsection, we describe differences between CTI and two closely related
metrics, \textit{National Chokepoint Potential (NCP)}~\cite{Leyba} 
and \textit{Hegemony}~\cite{Hegemony}.
\color{black}

\textbf{NCP}. Leyba {\em et al.} \cite{Leyba} identified topological bottlenecks, 
a framework that would also help in quantifying exposure to observation (as CTI aims to address), but with some
methodological differences, including: they identify transnational links
towards each country using delegation records, 
and they define bottleneck ASes as those serving the most paths 
(rather than IP addresses).
\color{black}
Further, both CTI and Leyba {\em et al.} \cite{Leyba} have as a goal
the identification of international inbound---and, in their case,
also outbound---\textit{chokepoints}
(\ie topological bottlenecks) in each country,
based on actual (CTI) or simulated (NCP) 
BGP paths towards each origin AS. However,
their work does not try to capture the fraction of the country's addresses
served by a transit provider, but rather the fraction of paths that a
border AS (\ie an AS which is registered to the same country as the origin, but which
has a neighbor that is registered to another country) may be able to intercept.
Our work is more narrowly focused on the specific case of a transit provider
serving traffic towards a transit-dominant country, taking into account the
address space of the direct or indirect customers.
Conceptually, weighting by paths
enhances the influence---or potential, in Leyba {\em et al.}'s
terminology---of ASes frequently serving a broad share of the country's
networks, whereas weighting by IPs yields higher influence
to ASes frequently serving a large fraction of the country's end hosts.

\color{black}
\textbf{Hegemony.} Our country-level transit influence metric is perhaps most similar to Hegemony~\cite{Hegemony}.
Both metrics aim to identify the transit ASes that are most prevalent
on paths towards origin ASes, weighted by the IP address space they serve. 
Hegemony can be applied either to the global AS-level graph, 
or to a ``Local graph: ... made only from AS paths with the same origin AS''~\cite{Hegemony}. 
The latter application is closest to CTI, as this analysis is 
limited to paths reaching a single origin AS; 
indeed, we use some of Hegemony local's filtering techniques
in our analysis (Sec.~\ref{sec:hegemony}). %
The applicability of (local-graph) Hegemony to the problem 
of revealing which transit ASes have observation capabilities
over traffic flowing towards a specific 
country---the issue addressed by CTI---is limited,
as Hegemony is a metric of centrality of transit ASes on 
a specific origin AS (not a country).  %

We build a country-level alternative metric
based on Hegemony~\cite{Hegemony} %
and compare CTI to it.
The reason for the comparison is to determine if CTI is too
aggressive in its filters, discarding too much input data.
For that purpose, we build a benchmark using Hegemony local,
a metric of centrality of any AS (including both transit providers and peers)
on paths towards a single origin.
Hegemony consists mostly
of a single filter on input BGP data, making it an appropriate benchmark.
This benchmark
was not trivial to build, as Hegemony local produces
a bilateral metric of influence between a transit AS and an origin
AS on the global topology. %
While Hegemony is concerned with extracting the most accurate estimate of centrality
on an existing graph, and not with estimating country-level inbound route diversity as CTI,
it is possible to build a metric that serves a similar purpose as CTI, which we
call \textit{country-level Hegemony} ($CLH$) as
\[ CLH(AS_t,C) \in [0,1] = \sum_{AS_o \in (C)} H(AS_t,AS_o) \cdot \dfrac{a^*(AS_o,C)}{A(C)},
\]
\noindent where $H(AS_t,AS_o)$ is the Hegemony score of $AS_t$ on $AS_o$ during the same
period\footnote{As Hegemony is published in 15-min intervals \cite{hegemonyapi}, we take the 5-day average score.} in March 2020 when we applied CTI,
(all the other terms have been previously introduced
in Eq. \ref{eq:countrytransit}).

We computed the absolute value of the difference between CTI and CLH
for each AS-coutry pair.
The output of
this analysis is (25th perc.,mean,median,75th perc.) =
0.00000, 0.00104, 0.00002, 0.00017, suggesting
that both metrics tend to agree about the country-level influence
of marginal ASes (the vast majority of AS-country pairs).
Therefore, we find no evidence that the heuristics of CTI %
introduce unnecessary noise to our analysis because,
on aggregate, a country-level alternative based on Hegemony---which
excludes considerably fewer BGP monitors than CTI does---tends
to agree with CTI's assessment.
The metrics do diverge on their assessment of ASes
that CTI has identified as influential
(CTI$\geq$0.1), with an avg. difference
between the metrics in those cases of 0.07.
\color{black}

\vspace{-0.2cm}

\section{Conclusions and Future Work}
\label{sec:conclusion}
In this work we tackled the issue of quantifying the exposure of a country's
traffic to observation or tampering by specific ASes. %
The Country-Level Transit Influence (CTI) metric we developed
aims to overcome several challenges with making such inferences using BGP
data. 
We apply this metric
in a set of---\color{black}potentially at-risk---countries
where transit provider-customer
relationships are still the dominant inbound
modality for international traffic; 
we identified these nations 
using both passive and active measurements.
\color{black} 
We applied CTI in these 75 countries %
and found that the median nation has 35\% of their IP addresses served by a single
transit AS.

\color{black}
In the future, we would like to develop measurement and analysis techniques
that can be applied to study the exposure of %
countries that are not primarily served by transit providers,
but rather by a dense mesh of bilateral and multilateral peering agreements,
\color{black} including those involving cloud providers and CDNs. \color{black}

\section*{Acknowledgements}
We thank our shepherd Amreesh Phokeer and the anonymous reviewers
for their insightful comments, and Amogh Dhamdhere and kc claffy 
for providing generous feedback. We are grateful 
to the network operators
who enabled our validation efforts.
This work was partly funded by the National Science Foundation (NSF),
Grant No. CNS 1705024. Author Gamero-Garrido was supported in part by 
the Microsoft Research Dissertation Grant (2019) and 
Northeastern University's Future Faculty Fellowship (2021).

\bibliographystyle{abbrv}
\bibliography{references}

\appendix

\section{BGP Monitor Location and CTI Process Diagram}
\label{app-bgp-mon}

\subsection{BGP Monitor Location}
We begin with the 685 monitors in RIPE and RouteViews. We discard (91)
monitors aggregated at multi-hop collectors and monitors that are not
full-feed, so we are left with 350 monitors in 209 ASes.  We determine
the location of each full-feed BGP monitor as follows.  First, we find
the locations of RouteViews and RIPE RIS BGP collectors. We build a
first set
of locations by finding RIPE Atlas probes co-located at Internet
Exchange Points (IXPs), by searching the list of peers for the IXP
name, and assign that probe to the country where the (single-location)
IXP is present, \eg BGP RRC01 -- LINX / LONAP, London, United Kingdom.
We confirm the BGP monitor location by running \texttt{ping}
measurements from RIPE Atlas probes hosted at the IXP to the BGP
monitor's IP address, and conclude that the BGP monitor is in the same
city as the IXP if the RTT is lower than 5~ms.
For the remaining BGP monitors we look for available RIPE Atlas probes
in the ASes that peer with the same BGP collector, and similarly run
\texttt{ping} measurements towards both the BGP monitor's IP address
and a RIPE Atlas probe located in the same city as the one listed for
the monitor. We conclude that the BGP monitor and RIPE Atlas
probe are in the same city if both sets of RTTs are under 5~ms.

We exclude 118 monitors at this stage because there is no available
RIPE Atlas probe hosted at the IXP (in the city where the monitor is
listed) nor at any of the other peers of the collector aggregating
announcements from the BGP monitor.  We discard remote peers from our
set, those that have \texttt{ping} RTTs higher than 30~ms from the RIPE Atlas
probe in the BGP monitor's listed city. For monitors with an RTT
between 5--30~ms, we infer them to be at the listed location if we get
confirmation using DNS records---\ie we find a geographical hint such
as a three-letter city or airport code, or the full name of the city,
using a reverse lookup with the BGP monitor's IP address---or a
matching country of the BGP monitor's \texttt{peer\_asn} record in the
RIPE RIS or RouteViews collector
list~\cite{routeviewslist,riperislist}.  Our final set $M$ has 214
monitors in 145 ASes and 19 countries.
We quantify the aggregate impact of all of our filters,
including the exclusion of certain BGP monitors per country,
in \S\ref{sec:hegcomparesec}, %
given an alternative
metric built using previous research \cite{Hegemony}.

\subsection{CTI Process Diagram}

\begin{figure}
    \centering
    \includegraphics[width=.99\textwidth]{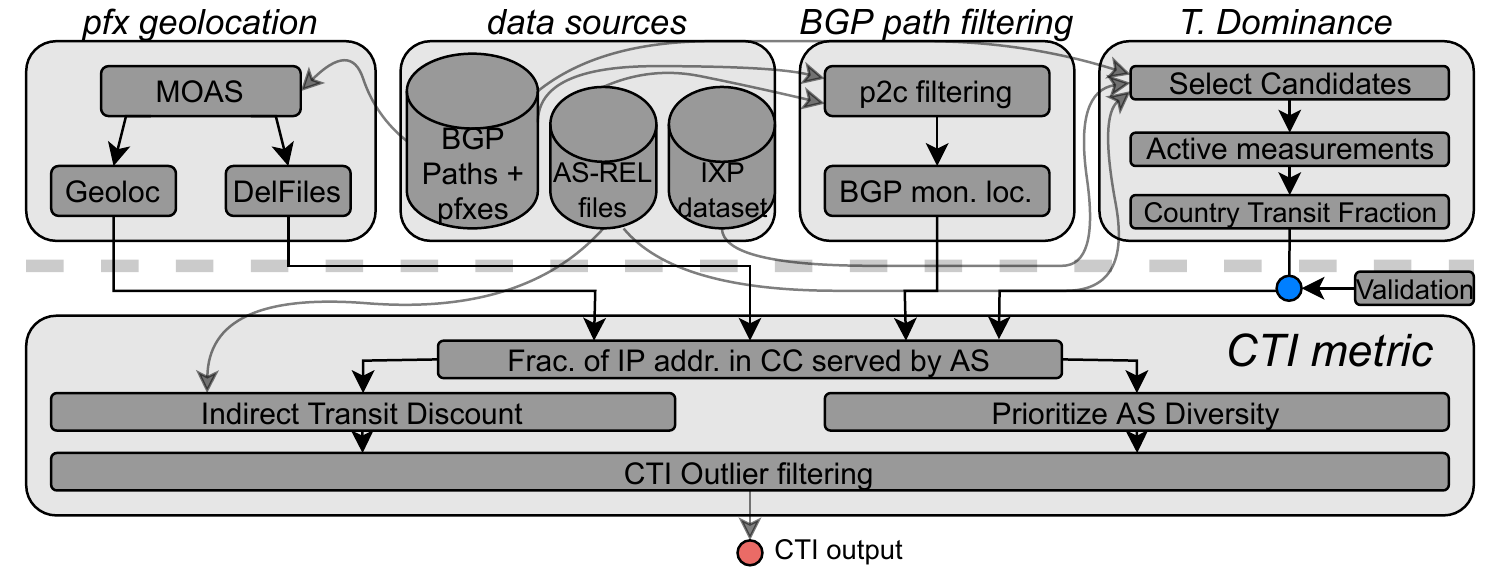}\\
    \caption{Process diagram showing input sources and analyses that produce our
        model of AS-level connectivity, the CTI metric.}\label{fig:ctidiagram}
\end{figure}

We show a process diagram of our methodology in Fig.~\ref{fig:ctidiagram}.
There, our transit-dominance country selection is shown in the top right 
corner, while the remaining blocks on the top row refer
to CTI inputs and preprocessing steps. Finally, the
bottom row shows the core components of the CTI metric.

\label{app:ctidiagram}

\end{document}